\documentclass[pre,showpacs,twocolumn,amsfonts,amsmath,nofootinbib]{revtex4}%
 \usepackage{pslatex}
 \usepackage{graphicx}

\begin{document}

 \title{A Quantitative Analysis of Measures of Quality in Science}
  \date{\today}
  \author{Sune Lehmann}
   \email{slj@imm.dtu.dk}
   \affiliation{Informatics and Mathematical Modeling, Technical University of
   Denmark, Building 321, DK-2800 Kgs.~Lyngby, Denmark.}
  \author{Andrew D.~Jackson}
   \affiliation{The Niels Bohr Institute, Blegdamsvej 17, DK-2100 K\o benhavn \O, Denmark.}
  \author{Benny E.~Lautrup}
   \affiliation{The Niels Bohr Institute, Blegdamsvej 17, DK-2100 K\o benhavn \O, Denmark.}

\newcommand{\be}{\begin{equation}}
\newcommand{\ee}{\end{equation}}
\newcommand{\bea}{\begin{eqnarray}}
\newcommand{\eea}{\end{eqnarray}}
\newcommand{\bdm}{\begin{displaymath}}
\newcommand{\edm}{\end{displaymath}}
\newcommand{\note}[1]{\textbf{(#1)\marginpar{!}}}

\begin{abstract}
Condensing the work of any academic scientist into a one-dimensional
measure of scientific quality is a difficult problem. Here, we
employ Bayesian statistics to analyze several different measures of
quality.  Specifically, we determine each measure's ability to
discriminate between scientific authors. Using scaling arguments, we
demonstrate that the best of these measures require approximately
$50$ papers to draw conclusions regarding long term scientific
performance with usefully small statistical uncertainties. Further,
the approach described here permits the value-free (i.e.,
statistical) comparison of scientists working in distinct areas of
science.
\end{abstract}


\maketitle

\section{Introduction}\label{sec:introduction}
It appears obvious that a fair and reliable quantification of the
`level of excellence' of individual scientists is a near-impossible
task \cite{adam:02,dellavalle:03,franck:99,adams:96,lehmann:06}.
Most scientists would agree on two qualitative observations:
\emph{(i)} It is better to publish a large number of articles than a
small number. \emph{(ii)} For any given paper, its citation
count---relative to citation habits in the field in which the paper
is published---provides a measure of its quality. It seems
reasonable to assume that the quality of a scientist is a function
of his or her full citation record\footnote{Citation data is, in
fact, publicly available for all academic scientists.}. The question
is whether this function can be determined and whether
quantitatively reliable rankings of individual scientists can be
constructed. A variety of `best' measures based on citation data
have been proposed in the literature and adopted in practice
\cite{raan:05,hirsch:05}. The specific merits claimed for these
various measures rely largely on intuitive arguments and value
judgments that are not amenable to quantitative investigation.
(Honest people can disagree, for example, on the relative merits of
publishing a single paper with $1000$ citations and publishing $10$
papers with $100$ citations each.)  The absence of quantitative
support for any given measure of quality based on citation data is
of concern since such data is now routinely considered in matters of
appointment and promotion which affect every working scientist.

Citation patterns became the target of scientific scrutiny in the
1960s as large citation databases became available through the work
of Eugene Garfield \cite{garfield:77} and other pioneers in the
field of bibliometrics. A surprisingly, large body of work on the
statistical analysis of citation data has been performed by
physicists. Relevant papers in this tradition include the pioneering
work of D.~J.~de~Solla Price, e.g.~\cite{price:65}, and, more
recently, \cite{redner:98,lehmann:03,hirsch:05,redner:05}. In
addition, physicists are a driving force in the emerging field of
complex networks. Citation networks represent one popular network
specimen in which papers correspond to nodes connected by references
(out-links) and citations (in-links). Citation networks have
frequently been used as an example of growing networks with
preferential attachment \cite{barabasi:99}. For reviews on this
extensive subject, see~\cite{albert:02,dorogovtsev:02,newman:03a}.
The aim of the present paper is to take such studies in a novel
direction by addressing the question of which one-dimensional
measure of citation data is best in a manner which is both
quantitative and free of value judgments. Given the remarks above,
the ability to answer this question depends on a careful definition
of the word `best'.

The primary purpose of analyzing and comparing the citation records
of individual scientists is to discriminate between them, i.e., to
assign some measure of quality and its associated uncertainty to
each scientist considered.  Whatever the intrinsic and value-based
merits of the measure, $m$, assigned to every author, it will be of
no practical value unless the corresponding uncertainty, $\delta m$
is sufficiently small. From this point of view, the best choice of
measure will be that which provides maximal discrimination between
scientists and hence the smallest value of $\delta m$. We will
demonstrate that the question of deciding which of several proposed
measures is most discriminating, and therefore `best', can be
addressed quantitatively using standard statistical methods.

Although the approach is straightforward, it is useful first to
describe it in general.  We begin by binning all authors by some
tentative measure, $m$, of the quality of their full citation
record. The probability that an author will lie in bin $\alpha$ is
denoted $p (\alpha )$. Similarly, we bin each paper according to the
total number of citations\footnote{We use the Greek alphabet when
binning with respect to to $m$ and the Roman alphabet for binning
citations.}. The full citation record for an author is simply the
set $\{n_i\}$, where $n_i$ is the number of his/her paper in
citation bin $i$. For each author bin, $\alpha$, we then empirically
construct the conditional probability distribution, $P(i | \alpha
)$, that a single paper by an author in this bin will lie in
citation bin $i$. These conditional probabilities are the central
ingredient in our analysis. They can be used to calculate the
probability, $P(\{n_i\}|\alpha)$, that any full citation record was
actually drawn at random on the conditional distribution,
$P(i|\alpha)$ appropriate for a fixed author bin, $\alpha$. Bayes'
theorem allows us to invert this probability to yield
\begin{equation}
P (\alpha | \{ n_i \} ) \sim P ( \{ n_i \} | \alpha ) \, p( \alpha )
\ , \label{intro1}
\end{equation}
where $P ( \alpha | \{ n_i \} )$ is the probability that the
citation record $\{ n_i \}$ was drawn at random from author bin
$\alpha$.  By considering the actual citation histories of authors
in bin $\beta$, we can thus construct the probability
$P(\alpha|\beta)$, that the citation record of an author initially
assigned to bin $\beta$ was drawn on the the distribution
appropriate for bin $\alpha$. In other words, we can determine the
probability that an author assigned to bin $\beta$ on the basis of
the tentative quality measure should actually be placed in bin
$\alpha$. This allows us to determine both the accuracy of the
initial author assignment its uncertainty in a purely statistical
fashion.

While a good choice of measure will assign each author to the
correct bin with high probability this will not always be the case.
Consider extreme cases in where we elect to bin authors on the basis
of measures unrelated to scientific quality, e.g., by hair/eye color
or alphabetically. For such measures $P( i | \alpha )$ and $P (  \{
n_i \} | \alpha  )$ will be independent of $\alpha$, and $P(\alpha |
\{n_i\})$ will become proportional to prior distribution
$p(\alpha)$. As a consequence, the proposed measure will have no
predictive power whatsoever. It is obvious, for example, that a
citation record provides no information of its author's hair/eye
color. The utility of a given measure (as indicated by the
statistical accuracy with which a value can be assigned to any given
author) will obviously be enhanced when the basic distributions $P
(i | \alpha )$ depend strongly on $\alpha$. These differences can be
formalized using the standard Kullback-Leibler divergence.  As we
shall see, there are significant variations in the predictive power
of various familiar measures of quality.

The organization of the paper is as follows.  Section~\ref{sec:data}
is devoted to a description of the data used in the analysis,
Section~\ref{sec:measures} introduces the various measures of
quality that we will consider. In Sections~\ref{sec:derivebayes} and
\ref{sec:weighing}, we provide a more detailed discussion of the
Bayesian methods adopted for the analysis of these measures and a
discussion of which of these measures is best in the sense described
above of providing the maximum discriminatory power. This will allow
us in Section~\ref{sec:scaling} to address to the question of how
many papers are required in order to make reliable estimates of a
given author's scientific quality; finally,
Section~\ref{sec:anomalies} discusses the origin of asymmetries in
some the measures. A discussion of the results and various
conclusions will be presented in Section~\ref{sec:conclusions}.

\section{Data}\label{sec:data}
The analysis in this paper is based on data from the
SPIRES\footnote{SPIRES is an acronym for 'Stanford Physics
Information REtrieval System'. The database is open and can be found
at http://www.slac.stanford.edu/spires/. Citations in SPIRES are
gathered only from the papers in the database that have references
entered electronically via eprints or journal articles, publications
such as monographs or conference proceedings are treated
inconsistently and therefore not included in this study.} database
of papers in high energy physics. Our data set consists of all
citable papers written by academic scientists from the theory
subfield, ultimo 2003. All citations to papers outside of SPIRES
were removed. In the context of this paper, we define an academic
scientist as someone who has published $25$ papers or more. This
definition is intended to include almost everyone with a permanent
academic position and exclude those who leave academia early in
their careers (and generally cease active journal publication) in
the interests of maintaining the homogeneity of the data sample. For
more see~\cite{lehmann:03a}, Chapters~3 and~4. The resulting data
set includes $6\,737$ authors and a total of $274\,470$ papers. The
actual number of papers is smaller than this since each multiple
author paper is counted once per co-author. The theory subfield is,
however, that part of high energy physics where this effect is least
pronounced. This is due to the relatively small number of co-authors
(typically $1-3$) per theoretical paper. In the case of the theory
subfield, this weighting of papers by the number of co-authors has
been shown to have negligible effects \cite{lehmann:03}.

The theory subsection of the SPIRES data has a power-law structure.
\begin{figure}
  \centering
  \includegraphics[width=.9\hsize]{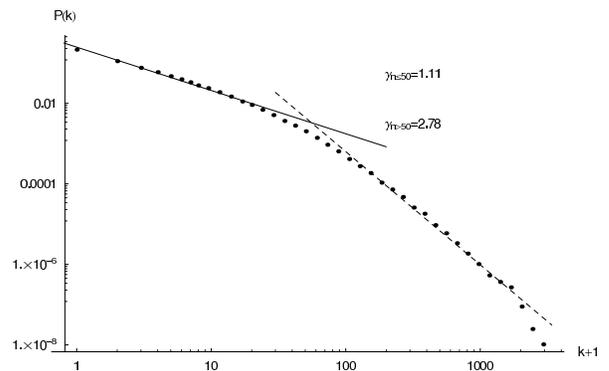}\\
  \caption{Logarithmically binned histogram of the citations counts of all papers by
  authors with more than $25$ publications in the theory
  subsection of SPIRES.
  The data is normalized and the axes are logarithmic.}\label{fig:totaldist}
\end{figure}
Specifically the probability that a paper will recieve $k$ citations
is approximately proportional to $(k+1)^{-\gamma}$ with $\gamma =
1.11$ for $k \le 50$ and $\gamma = 2.78$ for $k>50$. The transition
between these two power laws is found to be surprisingly
sharp~\cite{lehmann:03}.  These features of the global distribution
are also present in the conditional probabilities for sub-groups of
authors binned according to most measures of quality. In virtually
all cases, these conditional probabilities can also be described
accurately by separate power laws in each of two regions with a
relatively sharp transition between the regions. As one might
expect, authors with more citations are described by flatter
distributions (i.e., smaller values of $\gamma$) and a somewhat
higher transition point. Figure~\ref{fig:totaldist} displays the
total distribution of citations as a binned and normalized
histogram\footnote{Due to matters of visual presentation, the
binning used in this and the following figure here is different from
the binning used when constructing the $P(i|\alpha)$ used later in
the paper. The correct binning is described in
Appendix\ref{sec:explicit}}.

Studies performed on the first $25$, first $50$ and all papers for a
given value of $m$ show the absence of temporal correlations. It is
of interest to see this explicitly.
\begin{figure}
  \centering
  \includegraphics[width=.9\hsize]{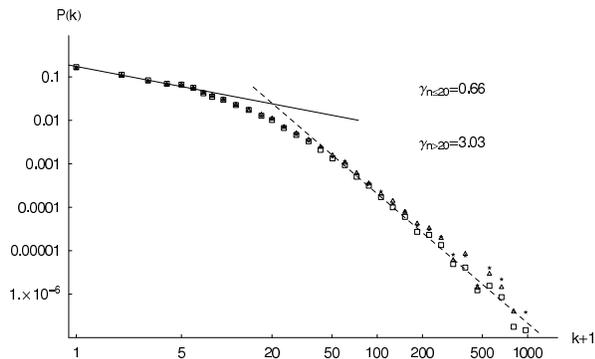}\\
  \caption{Logarithmically binned histogram of the citations in
  bin $6$ of the median measure. The $\triangle$ points show the
  citation distribution of the first $25$ papers by all authors.
  The points marked by {\Large $\star$} show the distribution of citations
  from the first $50$ papers by authors who have written more than
  $50$ papers. Finally, the $\Box$ data points show the distribution of all papers
  by all authors. The axes are logarithmic.}\label{fig:temporal}
\end{figure}
Consider the following example. In Figure~\ref{fig:temporal}, we
have plotted the distribution for bin $6$ of the median
measure\footnote{Since this plot is constructed from authors
assigned to bin 6, each paper is weighted by the number of its
authors present in this bin. Weighing papers by the number of
co-authors, however, does not significantly change the distribution
of citations \cite{lehmann:03}.}. There are $674$ authors in this
bin. Two thirds of these authors have written $50$ papers or more.
Only this subset is used when calculating the first $50$ papers
results. In this bin, the means for the total, first 25 and first 50
papers are $11.3$, $12.8$, and $12.9$ citations per paper,
respectively. The median of the distributions are $4$, $6$, and $6$.
The plot in Figure~\ref{fig:temporal} confirms these observations.
The remaining bins and the other measures yield similar results.

Note that Figure~\ref{fig:temporal} confirms the general
observations on the shapes of the conditional distributions made
above. Figure~\ref{fig:temporal} also shows two distinct power-laws.
Both of the power-laws in this bin are flatter than the ones found
in the total distribution and the transition point is lower than in
the total distribution from Figure\,\ref{fig:totaldist}.

\section{Measures of Scientific Excellence}\label{sec:measures}
Despite differing citation habits in different fields of science,
most scientists agree that the number of citations of a given paper
is the best objective measure of the quality of that paper. The
belief underlying the use of citations as a measure of quality is
that the number of citations to a paper provides an indication of
how often the content of that paper has been used in the work of
others\footnote{We realize that there are a number of problems
related to the use of citations as a proxy for quality. Papers may
be cited or not for reasons other than their high quality.  Geo-
and/or socio-political circumstances can keep works of high quality
out of the mainstream.  Credit for an important idea can be
attributed incorrectly.  Papers can be cited for historical rather
than scientific reasons.  Indeed, the very question of whether
authors actually read the papers they cite is not a simple one
\cite{simkin:03}. Nevertheless, we assume that correct citation
usage dominates the statistics.}. Note, however, the obvious fact
that citations can only be interpreted as a meaningful proxy of
quality relative to the citation habits of one's peers or, put
slightly differently, in the context of the citation habits of the
field in which the paper is published. In \cite{lehmann:03}, we have
shown that the theory subsection of SPIRES is indeed a very
homogeneous data set. In this sense, we will assume that the
citation count of a paper is a proxy of the intrinsic quality of
that paper.

The questions remain, however, of how to extract a measure of the
quality of an individual scientist from his citation record and how
fairly to project this record onto a scalar measure.  This question
is non-trivial because the probability, $p(k)$ of finding a
scientific paper with $k$ citations roughly follows an asymptotic
power-law distribution, see Figs.~\ref{fig:totaldist} and
\ref{fig:temporal}. This fact was documented for the SPIRES data in
Ref.~\cite{lehmann:03} and holds true in many other scientific
fields ~\cite{price:65,redner:98,newman:03a}. Thus, it is useful to
consider some of the properties of the distribution of citations for
all authors before discussing the various specific measures of
quality to be considered here.

Empirical evidence indicates that most citation distributions are
largely power-law distributed with $p(k) \sim k^{-\gamma}$. For
small values of $k$, $\gamma \approx 1$; for larger values, $2 <
\gamma < 3$. Although the average number of citations per paper is
well-defined, the asymptotic power-law tails of these distributions
cause their variance to be infinite\footnote{Diverging higher
moments of power-law distributions are discussed in the literature.
E.g.~\cite{newman:05a}.}.  When the variance is not defined (or very
large), the mean values of a finite sample fluctuate significantly
as a function of sample size. As a consequence, the average number
of citations, $\langle k \rangle$, in the citation record of a given
author (which is precisely a finite sample drawn from a power-law
probability distribution) is a potentially unreliable measure of the
quality of an author's citation record since the addition or removal
of a single highly cited paper can materially alter an author's
mean. Nevertheless, the mean of an author's citations is commonly
used as an intensive scalar measure of author quality.

The reservations just expressed about the use of mean citations per
paper apply with even greater force if one chooses to measure author
quality by the number of citations of each author's single most
highly cited paper, $k_{\rm max}$. Virtually all of the stabilizing
statistical power of the full citation record has been discarded,
and even greater fluctuations can be expected in this measure as the
sample size changes.  In spite of such statistical arguments, there
are reasons for considering the maximum cited paper as a measure of
quality.  It is perfectly tenable to claim that the author of a
single paper with $1000$ citations is of greater value to science
than the author of $10$ papers with $100$ citations each (even
though the latter is far less probable than the former).  In this
sense, the maximally cited paper might provide better discrimination
between authors of `high' and `highest' quality, and this measure
merits consideration.

Another simple and widely used measure of scientific excellence is
the average number of papers published by an author per year. This
would be a good measure if all papers were cited equally. As we have
just indicated, scientific papers are emphatically not cited
equally, and few scientists hold the view that all published papers
are created equal in quality and importance.  Indeed, roughly 50\%
of all papers in SPIRES are cited $\le 2$ times (including
self-citation).  This fact alone is sufficient to invalidate
publication rate as a measure of scientific excellence.  If all
papers were of equal merit, citation analysis would provide a
measure of industry rather than one of intrinsic quality.

In an attempt order to remedy this problem, \emph{Thomson
Scientific} (ISI) introduced the \textit{Impact Factor}\footnote{For
a full definition see http://scientific.thomson.com/.} which is
designed to be a ``measure of the frequency with which the `average
article' in a journal has been cited in a particular year or
period''\footnote{\emph{Ibid.}}. The Impact Factor can be used to
weight individual papers.  Unfortunately, citations to articles in a
given journal also obey power-law distributions~\cite{redner:05}.
This has two consequences.  First, the determination of the Impact
Factor is subject to the large fluctuations which are characteristic
of power-law distributions. Second, the tail of power-law
distributions displaces the mean citation to higher values of $k$ so
that the majority of papers have citation counts that are much
smaller than the mean. This fact is for example expressed in the
large difference between mean and median citations per paper. For
the total SPIRES data base, the median is $2$ citations per paper;
the mean is approximately $15$. Indeed, only $22\%$ of the papers in
SPIRES have a number of citations in excess of the mean,
cf.~\cite{lehmann:03}. Thus, the dominant role played by a
relatively small number of highly cited papers in determining the
Impact Factor implies that it is subject to relatively large
fluctuations and that it tends overestimate the level of scientific
excellence of high impact journals. This fact was directly verified
by Seglen~\cite{seglen:94}, who showed explicitly that the citation
rate for individual papers is uncorrelated to the impact factor of
the journal in which it was published.

An alternate way to measure excellence is to categorize each author
by the median number of citations of his papers, $k_{1/2}$. Clearly,
the median is far less sensitive to statistical fluctuations since
all papers play an equal role in determining its value. To
demonstrate the robustness of the median, it is useful to note that
the median of $\mathcal{N} = 2N+1$ random draws on \emph{any}
normalized probability distribution, $q(x)$, is normally distributed
in the limit $\mathcal{N}\to\infty$.  To this end we define the
integral of $q(x)$ as
\begin{equation}\label{eq:qintegral}
    Q(x) = \int^x q(x')dx'
\end{equation}
Evidently, $Q(x)$ grows monotonically from $0$ to $1$ independent of
$q(x)$.  The `median' of this sample is defined as that value of $x$
such that \emph{(i)}~one draw has the value $x$, \emph{(ii)}~$N$
draws have a value less than or equal to $x$, and \emph{(iii)}~$N$
draws have a value greater than or equal to $x$. The probability
that the median is at $x$ is now given as
\begin{equation}\label{eq:mediandist}
    P_{x_{1/2}}(x) = \frac{(2N+1)!}{1!N!N!}q(x)Q(x)^N[1-Q(x)]^N \ .
\end{equation}
For large $N$, the maximum of $P_{x_{1/2}}(x)$ occurs at $x=x_{1/2}$ where
$Q(x_{1/2})=1/2$.  Expanding $P_{x_{1/2}}(x)$ about its maximum value,
we see that
\begin{equation}\label{eq:mediandistapprox}
    P_{x_{1/2}}(x) = \frac{1}{\sqrt{2\pi\sigma^2}}
    \exp[-\frac{(x-x_{1/2})^2}{2\sigma^2}]\, ,\qquad
    \sigma^2 = \frac{1}{4 q(x_{1/2})^2 \mathcal{N}} \ .%
\end{equation}
A similar argument applies for every percentile. The statistical
stability of percentiles suggests that they are well-suited for
dealing with the power laws which characterize citation
distributions.

Recently, Hirsch \cite{hirsch:05} proposed a different measure, $h$,
intended to quantify scientific excellence. Hirsch's definition is
as follows:  ``A scientist has index $h$ if $h$ of his/her $N_p$
papers have at least $h$ citations each, and the other $(N_p-h)$
papers have fewer than $h$ citations each''\cite{hirsch:05}.  Unlike
the mean and the median, which are intensive measures largely
constant in time, $h$ is an extensive measure which grows throughout
a scientific career.  Hirsch assumes that $h$ grows approximately
linearly with an author's professional age, defined as the time
between the publication dates of the first and last paper.
Unfortunately, this does not lead to an intensive measure. Consider,
for example, the case of authors with large time gaps between
publications, or the case of authors whose citation data are
recorded in disjoint databases. A properly intensive measure can be
obtained by dividing an author's $h$-index by the number of his/her
total publications. We will consider both approaches below.

The $h$-index represents an attempt to strike a balance between
productivity and quality and to escape the tyranny of power law
distributions which place strong weight on a relatively small number
of highly cited papers. The problem is that Hirsch assumes an
equality between incommensurable quantities. An author's papers are
listed in order of decreasing citations with paper $i$ having $C(i)$
citations. Hirsch's measure is determined by the equality, $h =
C(h)$, which posits an equality between two quantities with no
evident logical connection. While it might be reasonable to assume
that $h^{\gamma} \sim C(h)$, there is no reason to assume that
$\gamma$ and the constant of proportionality are both $1$.

We will also include one intentionally nonsensical choice in the
following analysis of the various proposed measures of author
quality.  Specifically, we will consider what happens when authors
are binned alphabetically.  In the absence of historical
information, it is clear that an author's citation record should
provide us with no information regarding the author's name. Binning
authors in alphabetic order should thus fail any statistical test of
utility and will provide a useful calibration of the methods
adopted. The measures of quality described in this section are the
ones we will consider in the remainder of this paper.

\section{A Bayesian Analysis of Citation Data}\label{sec:derivebayes}
The rationale behind all citation analyses lies in the fact that
citation data is strongly correlated such that a `good' scientist
has a far higher probability of writing a good (i.e., highly cited)
paper than a `poor' scientist.  Such correlations are clearly
present in SPIRES ~\cite{lehmann:03,lehmann:05}. We thus categorize
each author by some tentative quality index based on their total
citation record. Once assigned, we can empirically construct the
prior distribution, $p(\alpha)$, that an author is in author bin
$\alpha$ and the probability $P(N|\alpha)$ that an author in bin
$\alpha$ has a total of $N$ publications. We also construct the
conditional probability $P ( i | \alpha )$ that a paper written by
an author in bin $\alpha$ will lie in citation bin $i$. As we have
seen earlier, studies performed on the first $25$, first $50$ and
all papers of authors in a given bin reveal no signs of additional
temporal correlations in the lifetime citation distributions of
individual authors. In performing this construction, we have elected
to bin authors in deciles. We bin papers into $L$ bins according to
the number of citations. The binning of papers is approximately
logarithmic (see Appendix A). We have confirmed that the results
stated below are largely independent of the bin-sizes chosen.

We now wish to calculate the probability, $P ( \{ n_i \} | \alpha
)$, that an author in bin $\alpha$ will have the full (binned)
citation record $\{ n_i \}$.  In order to perform this calculation,
we assume that the various counts $n_i$ are obtained from $N$ {\em
independent\/} random draws on the appropriate distribution, $P( i |
\alpha )$.  Thus,
\begin{equation}\label{eq:conditionalindependence}
    P(\{n_i \}|\alpha ) = P(N|\alpha)N!\prod_{i=1}^L \frac{P(i|\alpha )^{n_i}}{(n_i)!} \ .
\end{equation}
Although large scale temporal correlations are known to be absent,
transient correlations are possible. For example, one particularly
well-cited paper could lead to an increased probability of high
citations for its immediate successor(s). It is difficult to
demonstrate their presence or absence, but the results of following
section will provide a posteriori evidence that such correlations,
if present, are not important.

\begin{figure*}
\centering
 \begin{tabular}{cccc}
  \emph{(a)} First initial&\emph{(b)} Papers/year &\emph{(c)} Hirsch (age) & \emph{(d)} Hirsch (papers)\\
  \includegraphics[width=.24\hsize]{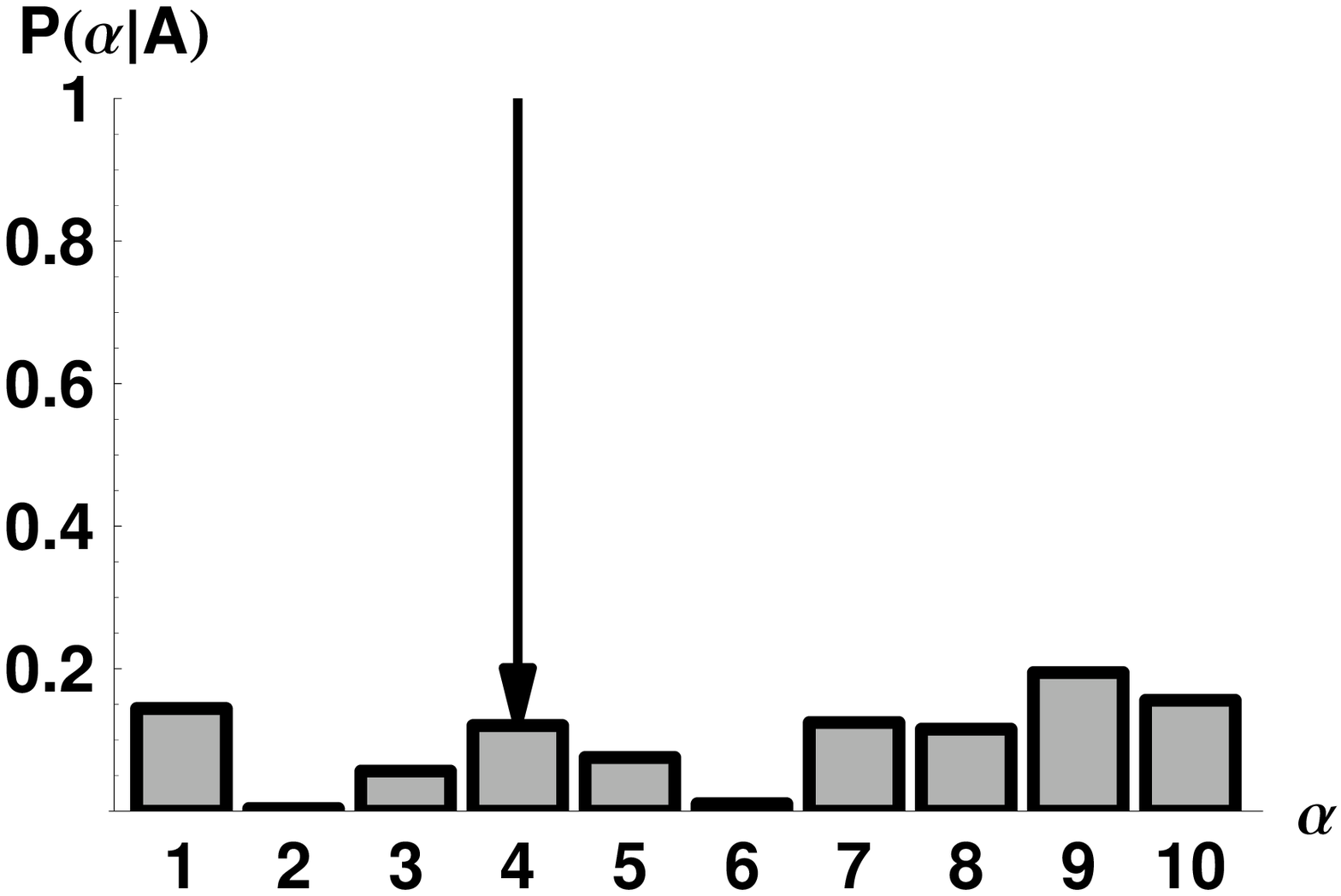} &%
  \includegraphics[width=.24\hsize]{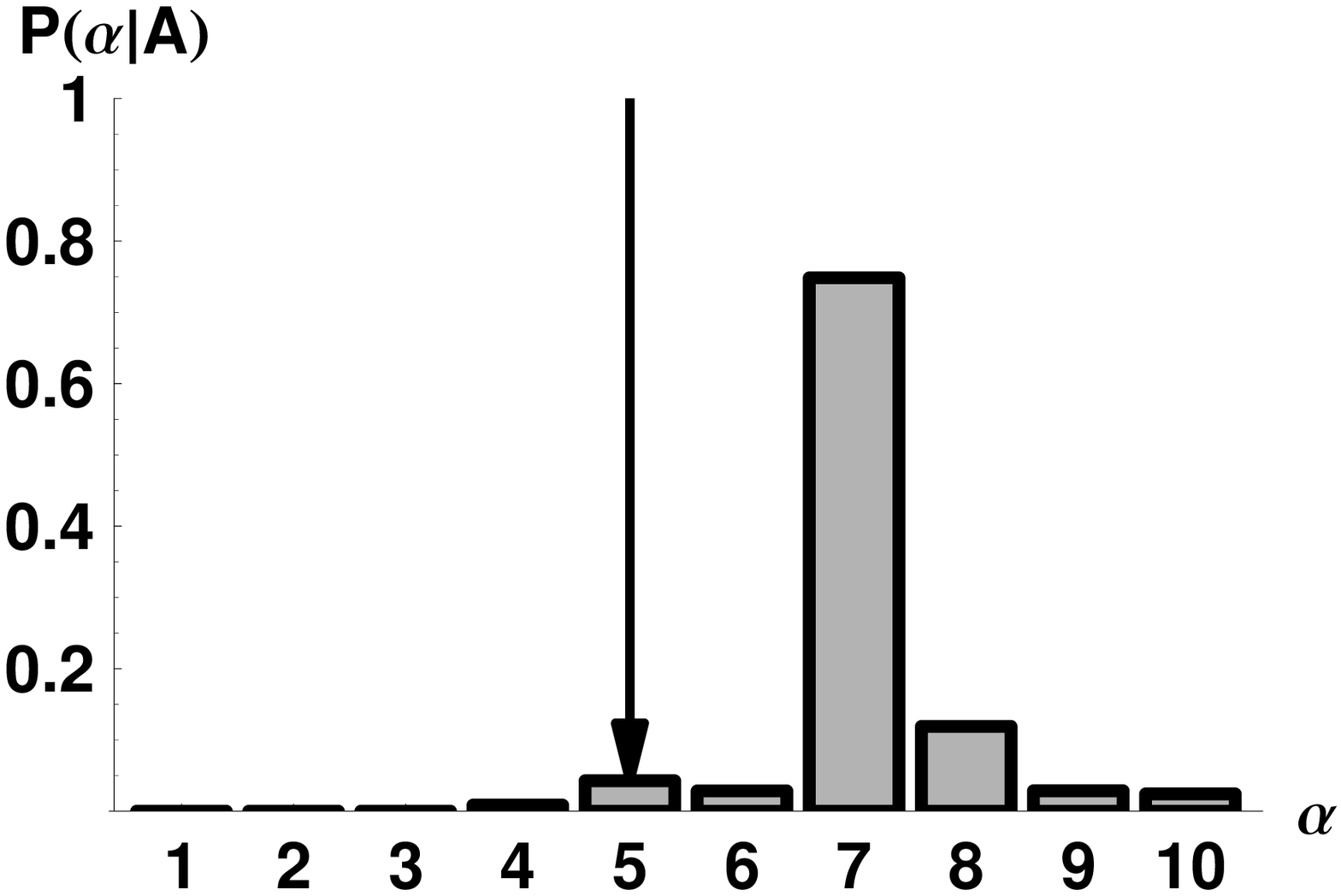}& %
  \includegraphics[width=.24\hsize]{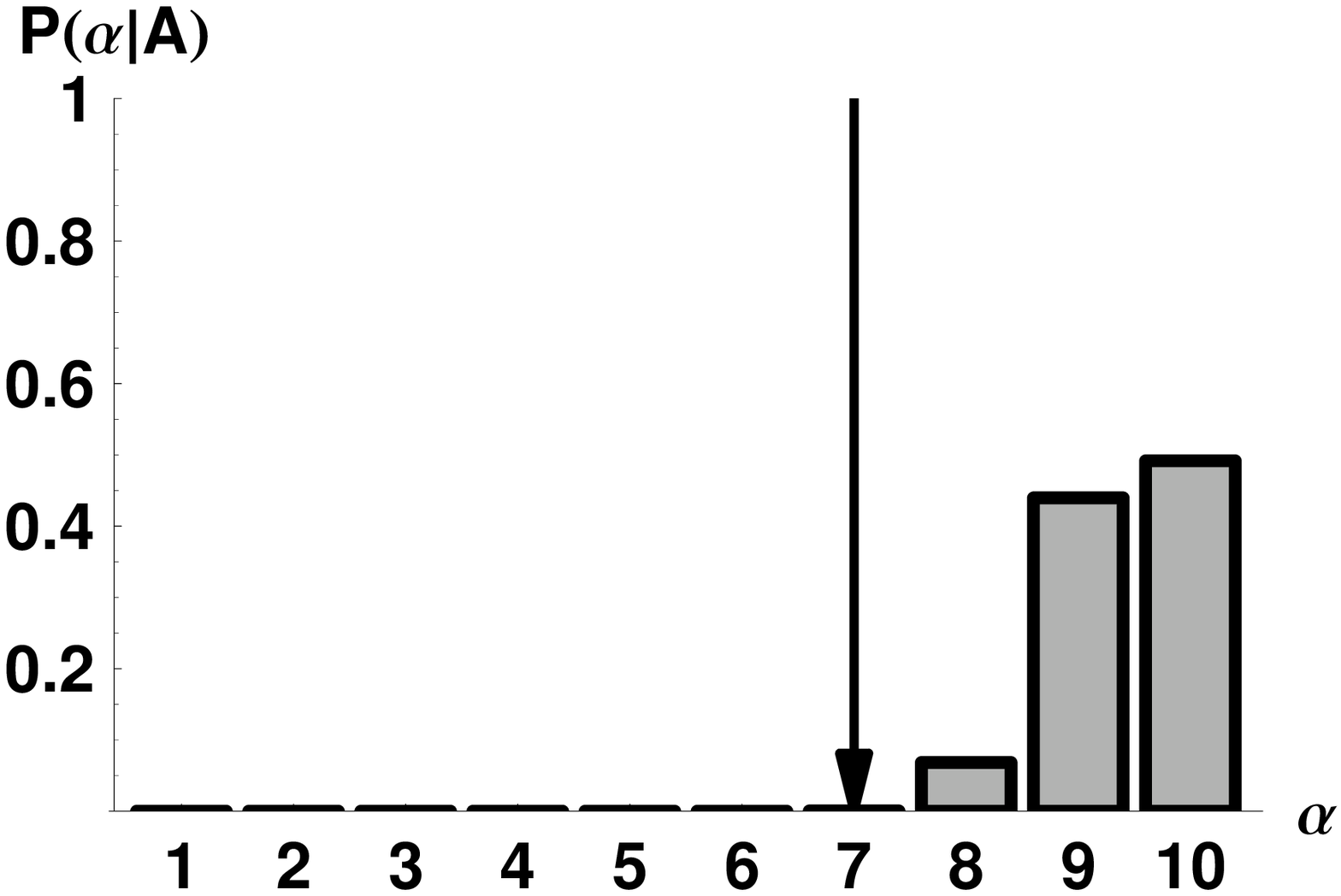}& %
  \includegraphics[width=.24\hsize]{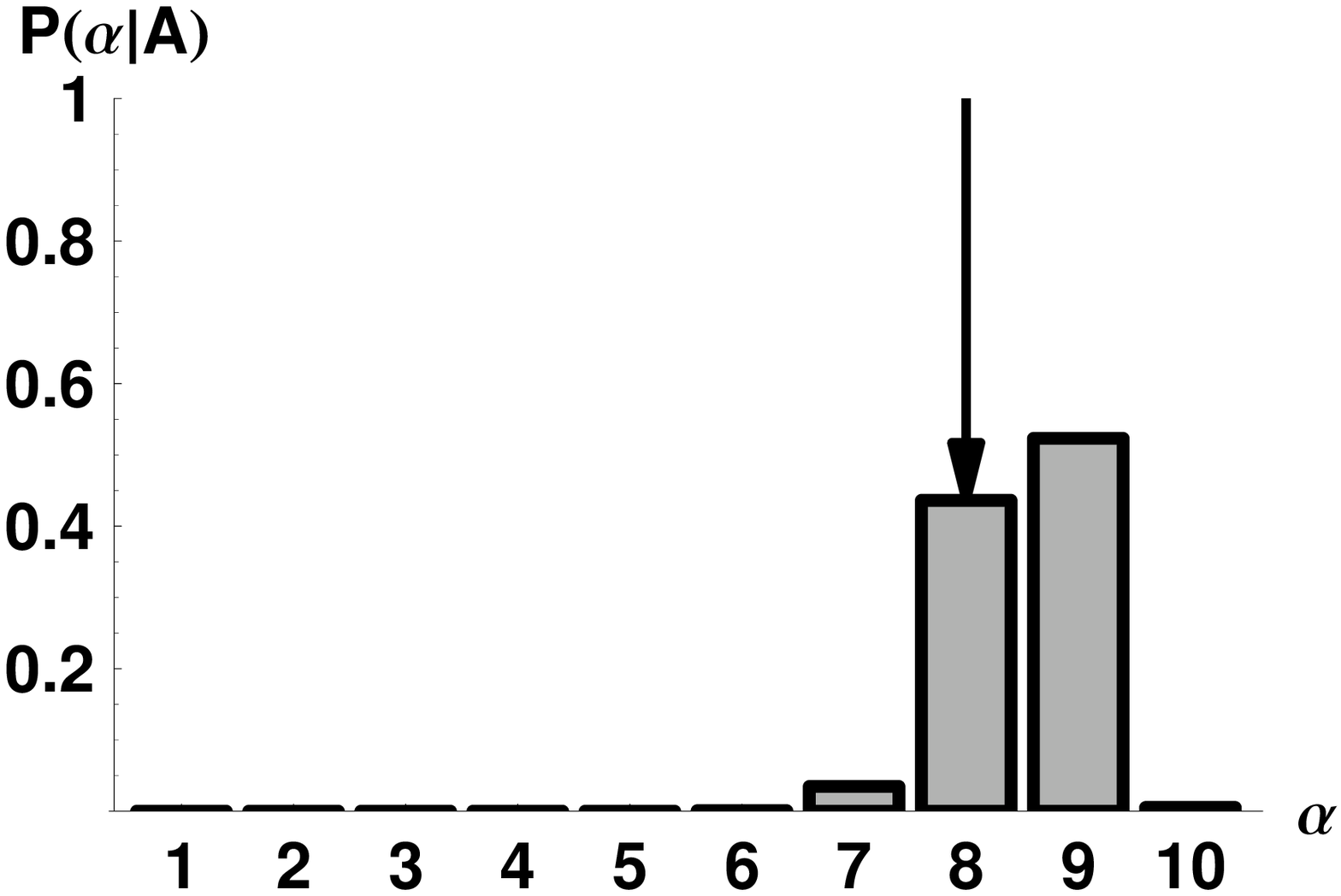}\\%
  \emph{(e)} Max &\emph{(f)} Mean &\emph{(g)} Median & \emph{(h)} 65th percentile\\
  \includegraphics[width=.24\hsize]{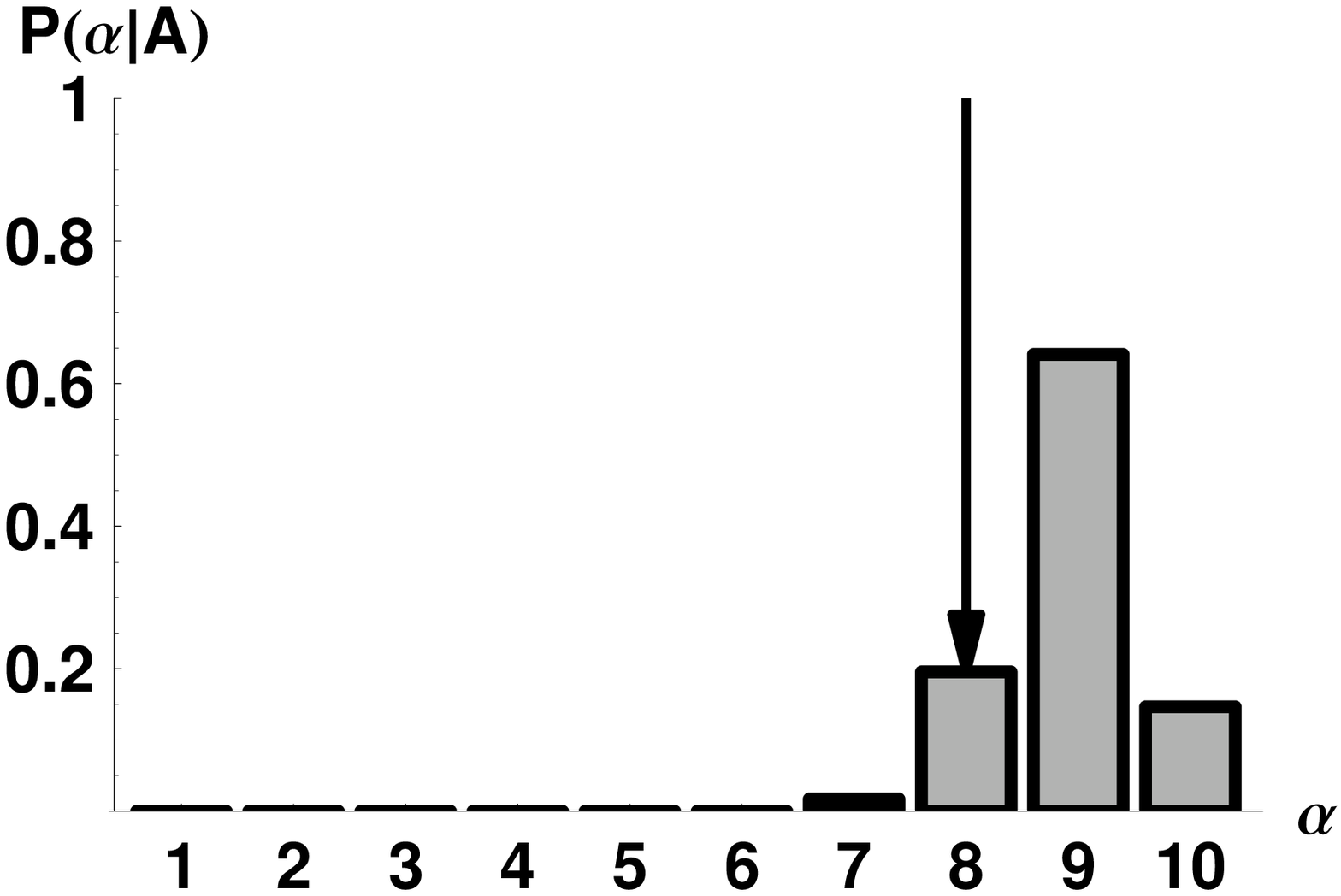} &  %
  \includegraphics[width=.24\hsize]{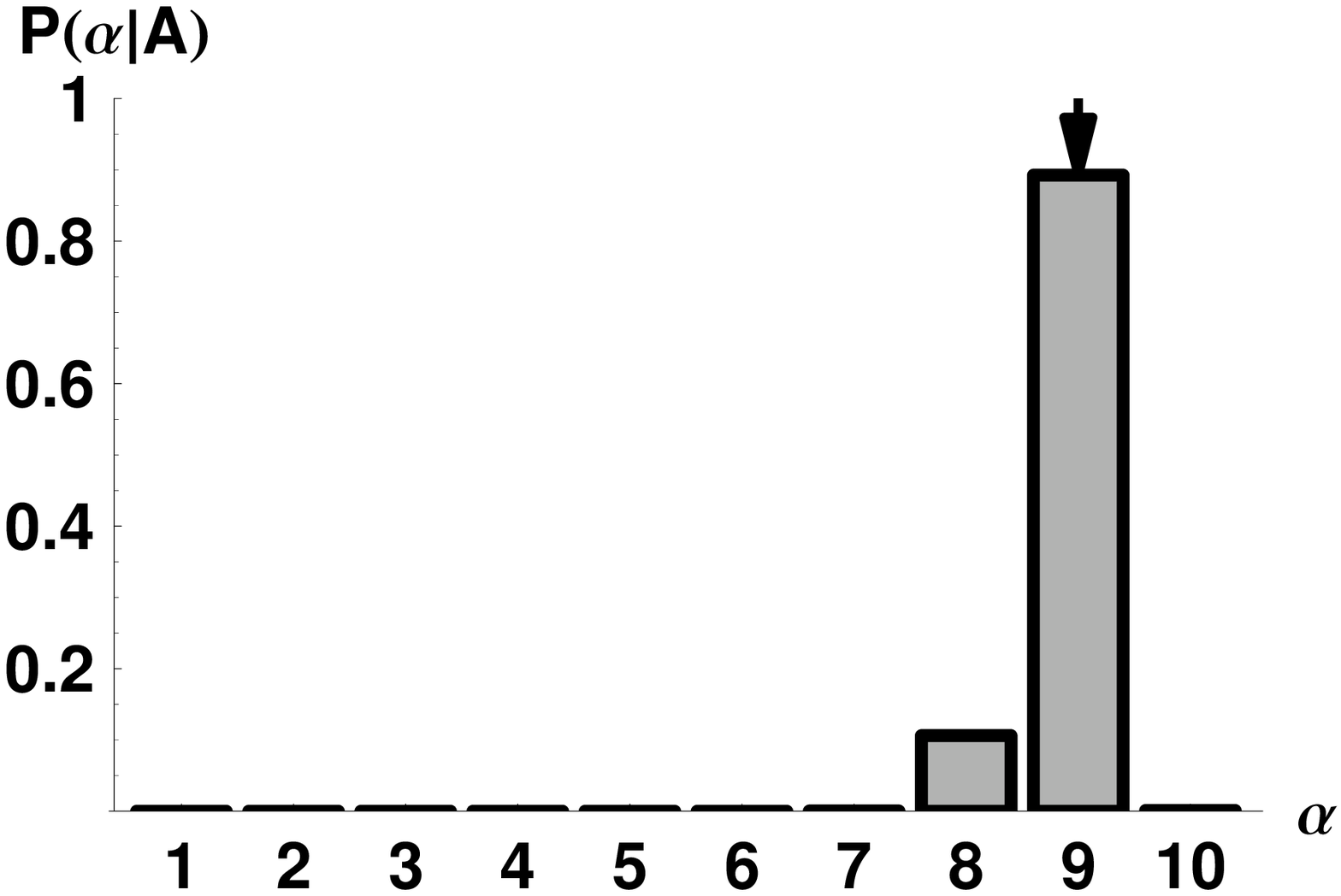} &%
  \includegraphics[width=.24\hsize]{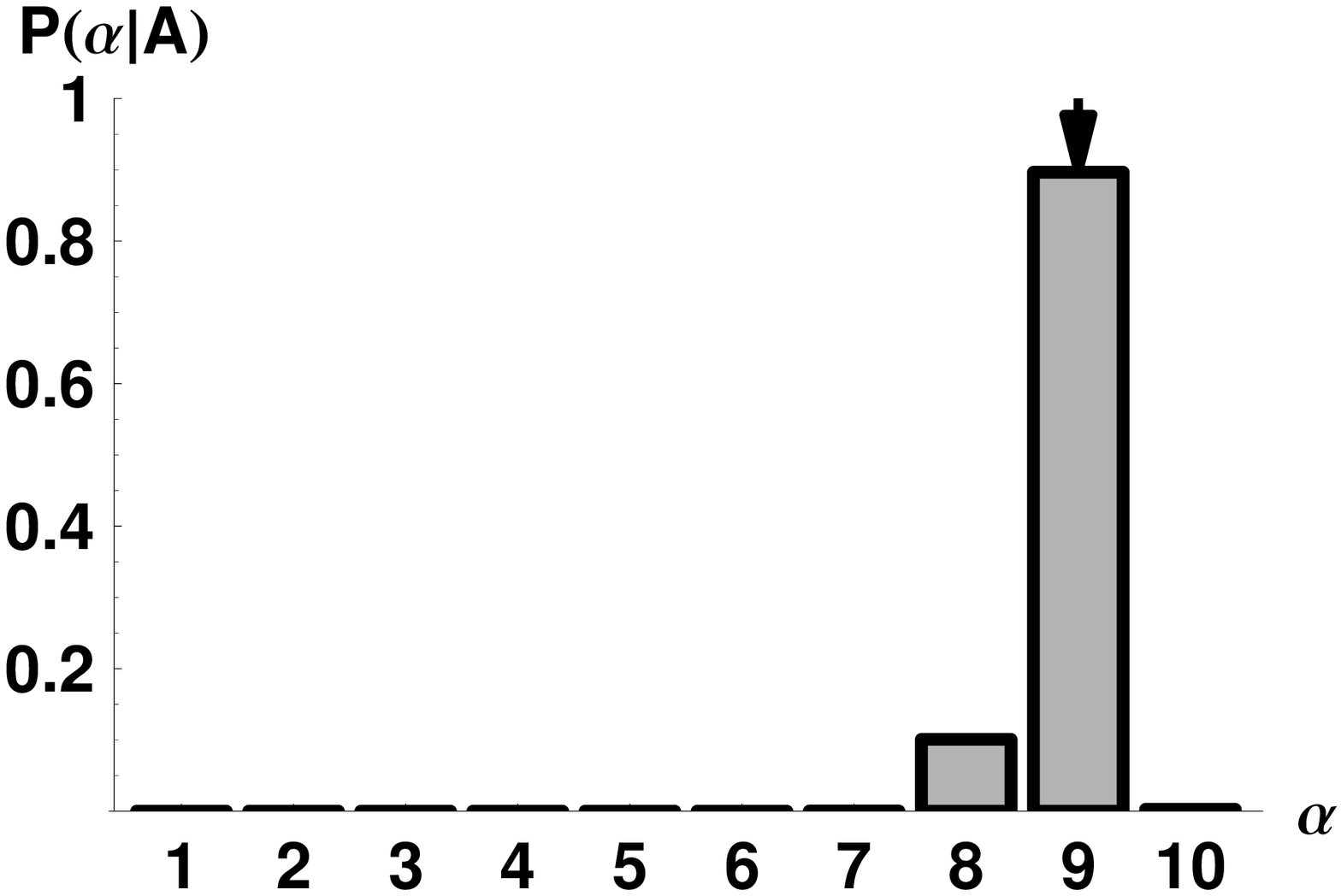}&%
  \includegraphics[width=.24\hsize]{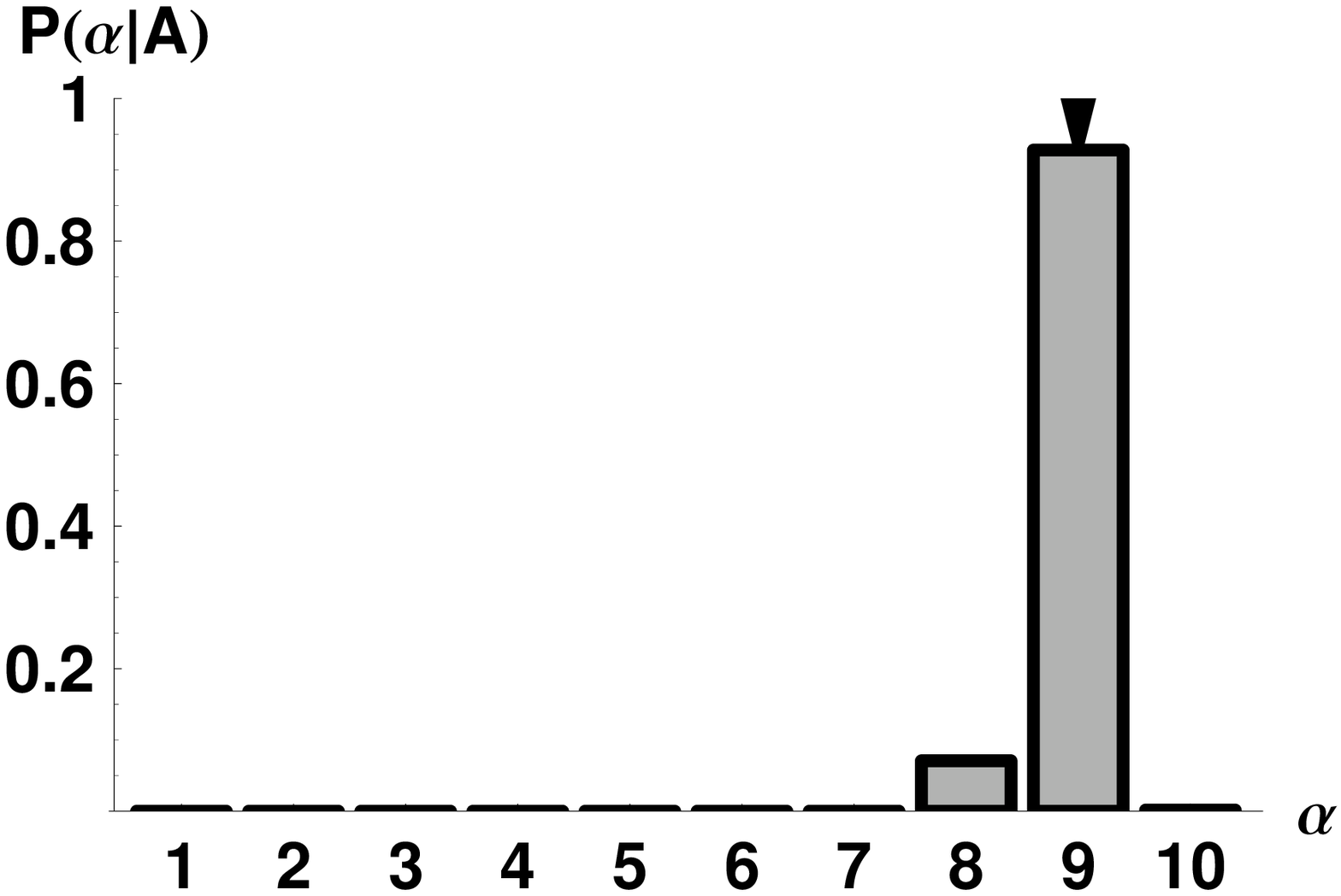}
 \end{tabular}
\caption{A single author example. We analyze the citation record of
author $A$ with respect to the eight different measures defined in
the text. Author $A$ has written a total of $88$ papers. The mean of
this citation record is $26$ citations per paper, the median is $13$
citations, the $h$-index is $29$, the maximally cited paper has
$187$ citations, and papers have been published at the average rate
of $2.5$ papers per year. The various panels show the probability
that author $A$ belongs to each of the ten deciles given on the
corresponding measure; the vertical arrow displays the initial
assignment. Panel \emph{(a)} displays $P(\text{first initial}|A)$
\emph{(b)} shows $P(\text{papers per year}|A)$, \emph{(c)} shows
$P(h/T|A)$, \emph{(d)} shows $P(h/N|A)$, panel \emph{(e)} shows
$P(k_{\rm max}|A)$, panel \emph{(f)} displays $P(\langle k
\rangle|A)$, \emph{(g)} shows $P(k_{1/2}|A)$ , and finally
\emph{(h)} shows $P(k_{.65}|A)$. } \label{fig:example}
\end{figure*}

We can now invert the probability $P(\{n_j\}|\alpha)$ using Bayes'
theorem to obtain
\begin{eqnarray}
    P(\alpha |\{n_i\}) &=& \frac{P(\{n_i\}|\alpha) \, p(\alpha )}{P(\{n_i\})}\nonumber\\%
                    &=& \frac{p( \alpha ) P(N|\alpha)\prod_{j} P(j|\alpha)^{n_j}}{\sum_{\beta}
                      p( \beta ) P(N|\beta)\, \prod_{k} P(k|\beta)^{n_{k}}} \ ,
\label{eq:bayes}%
\end{eqnarray}
where we have inserted Eq.~(\ref{eq:conditionalindependence}) and
used marginalization to obtain the normalization. The combinatoric
factors cancel.  The quantity $P ( \alpha | \{ n_i \} )$, which
represents the probability that an author with binned citation
record $\{ n_i \}$ is in author bin $\alpha$. It can be used in two
ways---each of which is interesting.

For any measure chosen Eq.~(\ref{eq:bayes}) provides us with the
probability that an author lies in author bin $\alpha$. While the
value of any measure (such as the mean number of citations per
paper) can be calculated directly, the calculated values of $P (
\alpha | \{ n_i \} )$ provide far more detailed and more reliable
information using {\em all\/} statistical information contained in
the data.  The large fluctuations which can be encountered in
identifying authors by their mean citation rate or by their
maximally cited paper are reduced.  Further, by providing us with
values of $P ( \alpha | \{ n_i \} )$ for all $\alpha$, we obtain a
statistically trustworthy gauge of whether the resulting
uncertainties in $\alpha$ are sufficiently small for the measure
under consideration to be a reliable indicator of author quality. In
short, Eq.~(\ref{eq:bayes}) provides us with a measure of an
author's ranking independent of the total number papers currently
published, and with information which allows us to assess the
reliability of this determination. The accuracy of the resulting
value of $\alpha$ increases dramatically with the total number of
published papers.  We will return to this point in Section V.

Fig.~\ref{fig:example} shows the probabilities $P(\alpha|\{n_i\})$
that $A$ will lie in each of the decile bins using the measures
discussed in section II. These measures include: \emph{(a)} the
first initial of the author's name, \emph{(b)} the average yearly
output of papers, \emph{(c)} Hirsch's~$h$ normalized by the author's
professional age $T$, \emph{(d)} the $h$-index normalized by the
number of published papers, \emph{(e)} the citation count of the
single most cited paper, \emph{(f)} the mean number of citations per
paper, \emph{(g)} the median number (50th percentile) of citations
per paper, and \emph{(h)} a 65th percentile measure. It is clear
from the figure that there are significant differences, both in the
accuracy of of the initial assignments and, more importantly, in the
corresponding uncertainties.  Large uncertainties are due to the
fact that the conditional probabilities, $P ( i | \alpha )$ are
largely independent of $\alpha$. Such independence is to be expected
in the case of the alphabetic binning of authors, where the
inability of the citation record to identify the first initial of
author $A$'s name is hardly surprising.  The figure also suggests
that the number of papers published per year is not reliable.
Initial assignments of author $A$  based on mean, median, 65th
percentile, and maximum citations all appear to provide an accurate
reflection of his full citation record with a satisfactorily small
uncertainty. Hirsch's measures falls somewhere between the best and
worst choice of measures.

Given the large variations in the accuracy and confidence of
decile assignments as a function of the measure selected, it is of interest to
investigate in greater detail the question of which of these measures is best.
We address this question in the next section.

\section{Weighing the Measures}\label{sec:weighing}
\begin{figure*}
\centering
\begin{tabular}{cccc}
  \emph{(a)} First Initial &\emph{(b)} Papers/year &\emph{(c)} Hirsch (age) & \emph{(d)} Hirsch (papers)\\
  \includegraphics[width=.24\hsize]{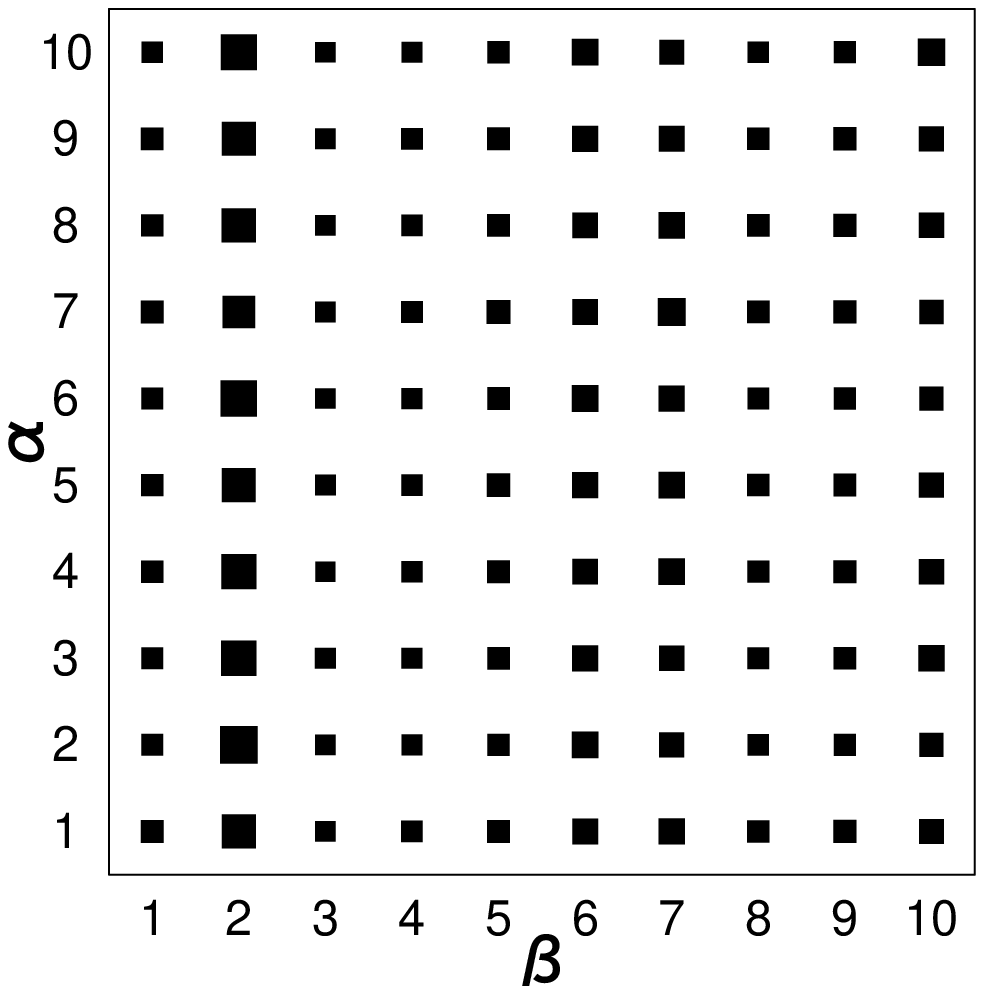} &%
  \includegraphics[width=.24\hsize]{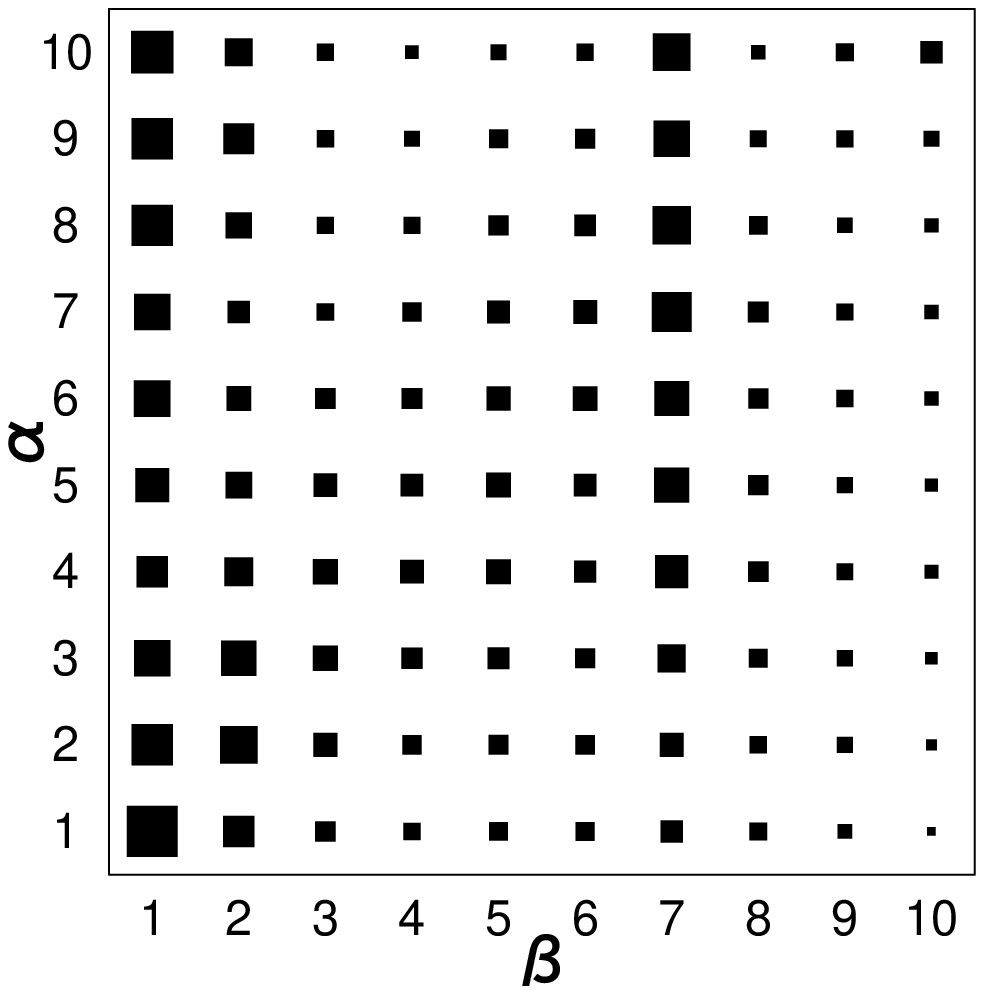} &%
  \includegraphics[width=.24\hsize]{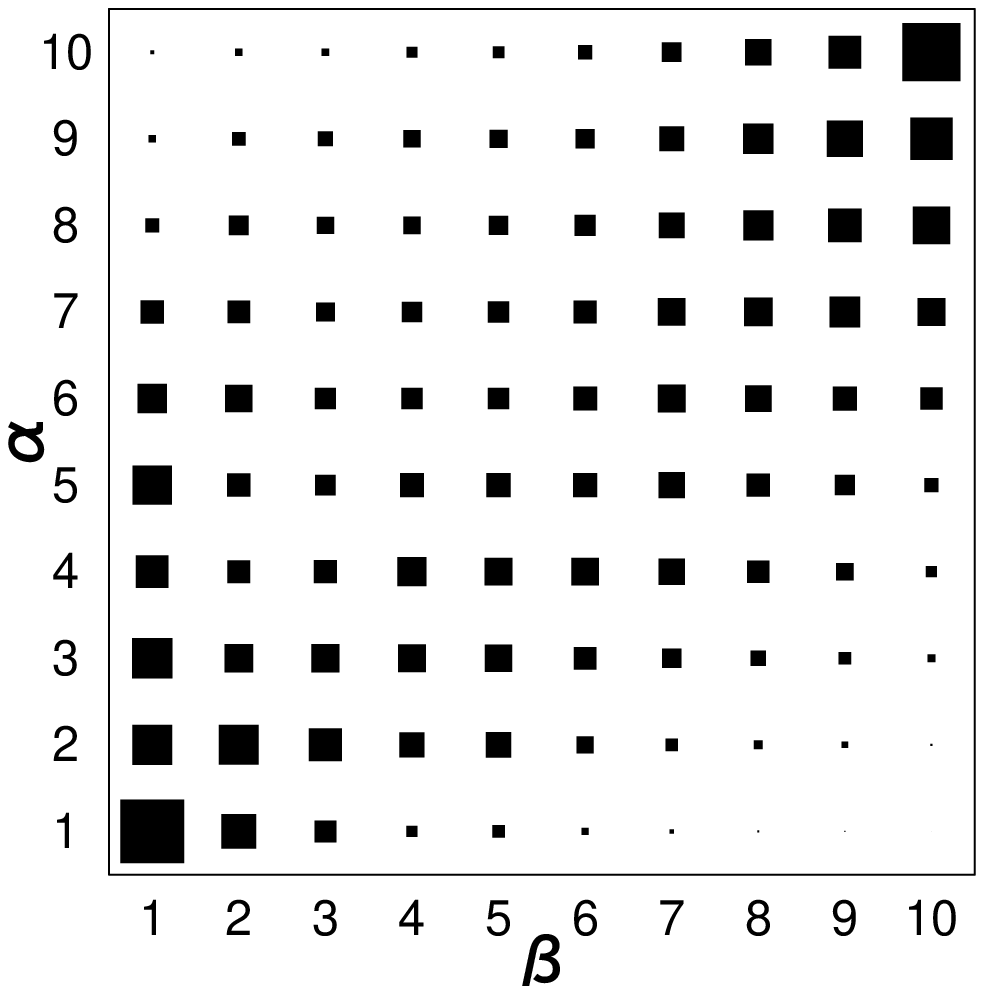} &%
  \includegraphics[width=.24\hsize]{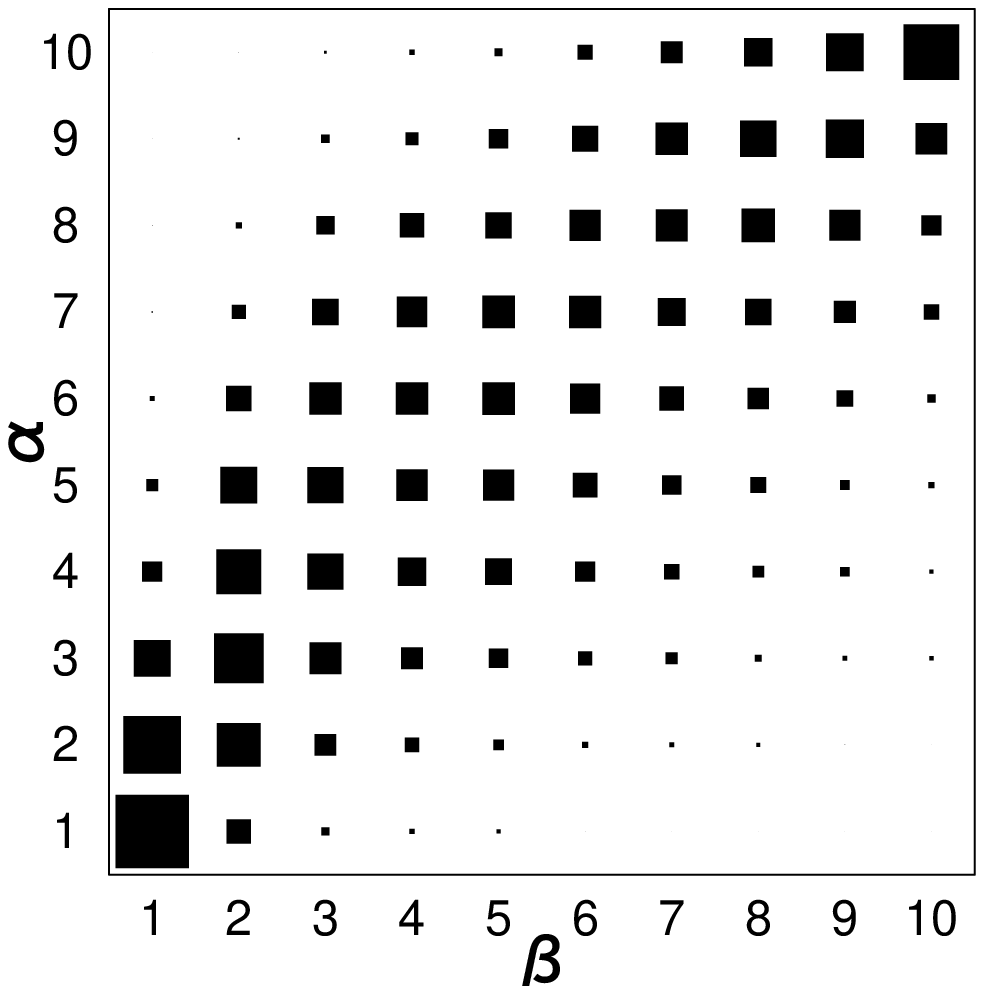}\\
  \emph{(e)} Max &\emph{(f)} Mean &\emph{(g)} Median & \emph{(h)} 65th percentile\\
  \includegraphics[width=.24\hsize]{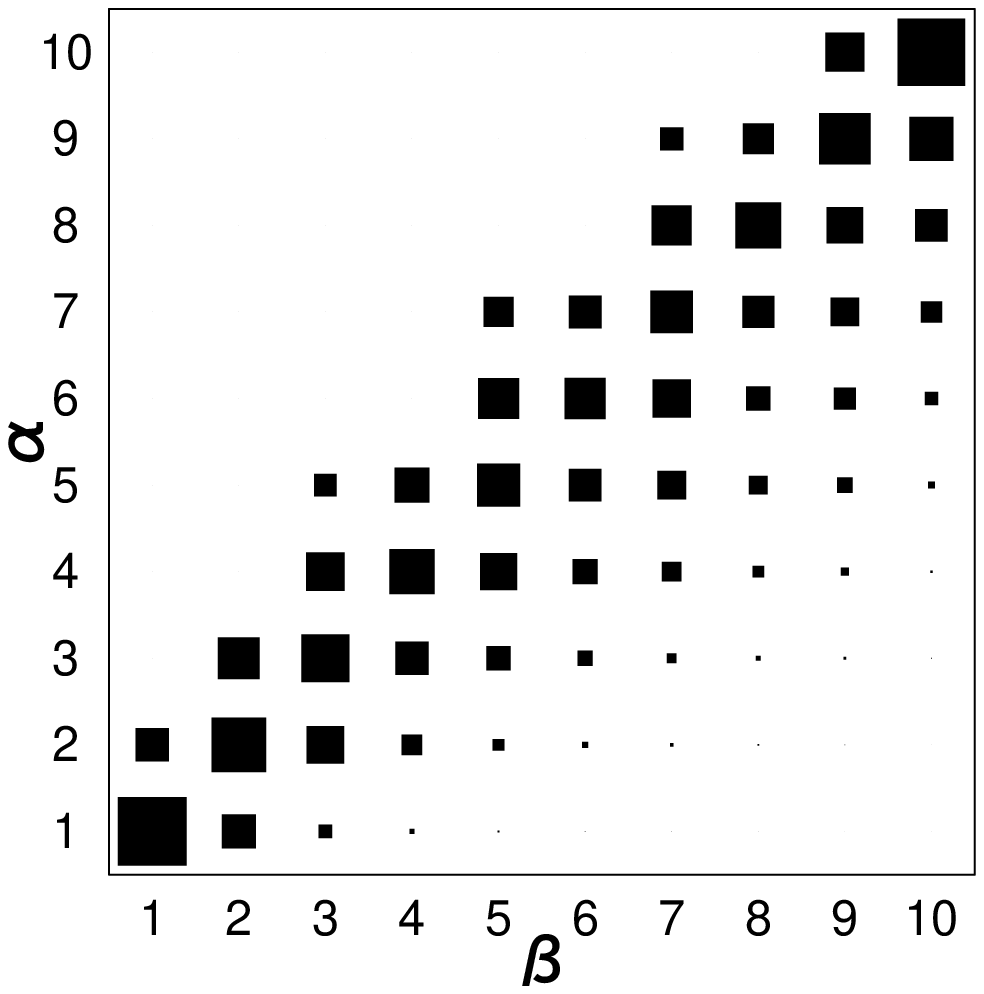} &%
  \includegraphics[width=.24\hsize]{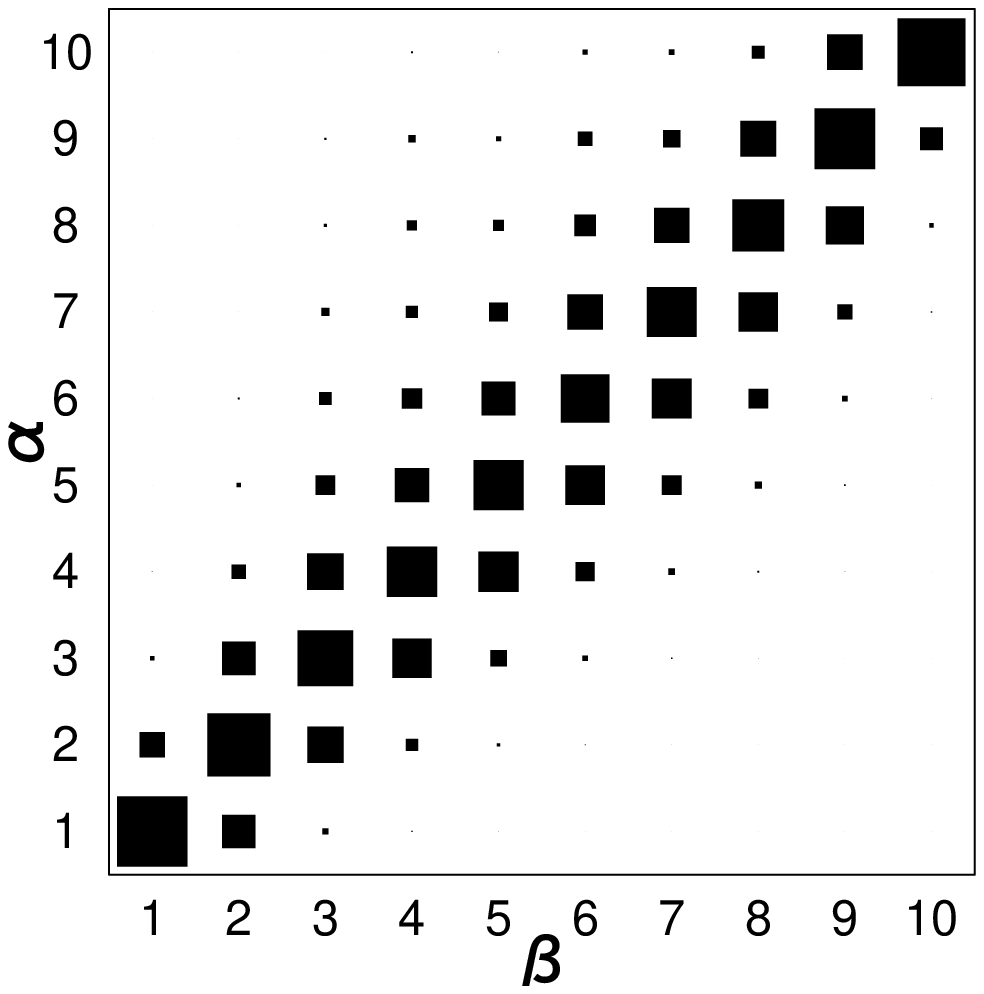} &%
  \includegraphics[width=.24\hsize]{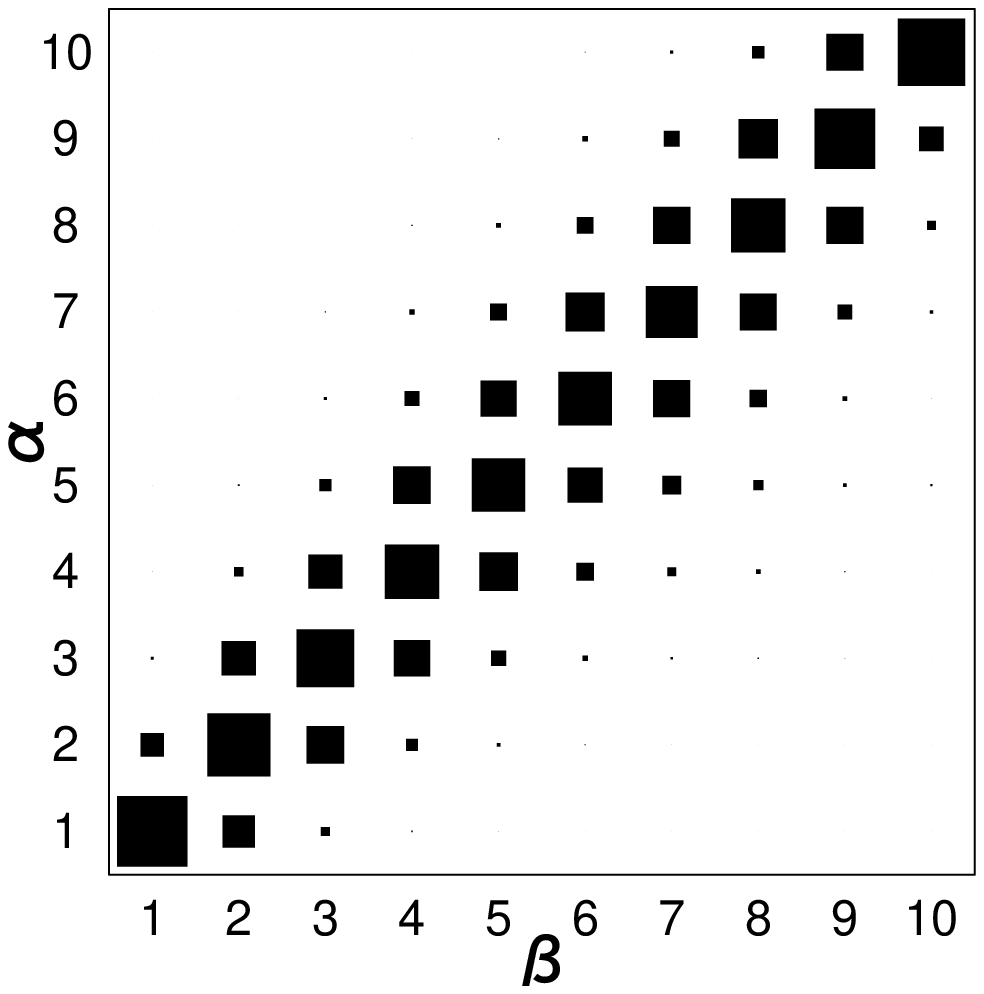}&%
  \includegraphics[width=.24\hsize]{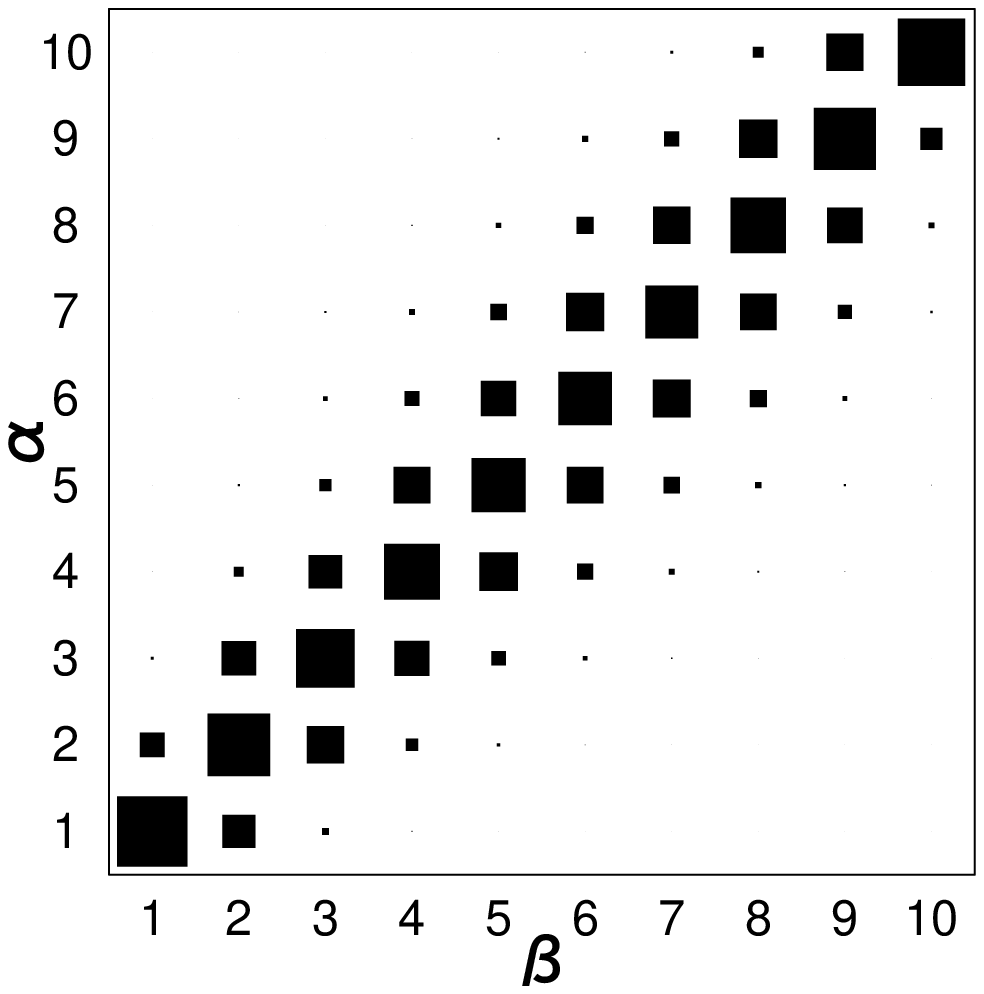}
\end{tabular}
\caption{Eight different measures. Each horizontal row shows the
average probabilities (proportional to the areas of the squares)
that authors initially assigned to decile bin $\alpha$ are predicted
to belong in bin $\beta$. Panels as in Fig.~3.} \label{fig:avePnms}
\end{figure*}

In order to obtain a more graphic representation of the quality of a
given measure, we calculate the probability, $P (\beta | \alpha )$,
that an author initially assigned to bin $\alpha$ is predicted to
lie in bin $\beta$. In practice, we determine $P ( \beta | \alpha )$
as the average of the probability distributions $P ( \beta | \{ n_i
\} )$ for each author in bin $\alpha$. The results are shown
`stacked' in Fig.~\ref{fig:avePnms} for the various measures
considered. Here, row $\alpha$ shows the (average) probabilities
that an author initially assigned to bin $\alpha$ belongs in decile
bin $\beta$. This probability is proportional to the area of the
corresponding squares.  Obviously, a perfect measure would place all
of the weight in the diagonal entries of these plots. Weights should
be centered about the diagonal for an accurate identification of
author quality and the certainty of this identification grows as
weight accumulates in the diagonal boxes.  Note that an assignment
of a decile based on Eq.~(\ref{eq:bayes}) is likely to be more
reliable than the value of the initial assignment since the former
is based on all information contained in the citation record.

Figure~\ref{fig:avePnms} emphasizes that `first initial' and
`publications per year' are not reliable measures. The $h$-index
normalized by professional age performs poorly; when normalized by
number of papers, the trend towards the diagonal is enhanced. We
note the appearance of vertical bars in each figure in the top row.
This feature is explained in Appendix \ref{sec:anomalies}. All four
measures in the bottom row perform fairly well. The initial
assignment of the $k_{\rm max}$ measure always underestimates an
author's correct bin. This is not an accident and merits comment.
Specifically, if an author {\em has\/} produced a single paper with
citations in excess of the values contained in bin $\alpha$, the
probability that he will lie in this bin, as calculated with
Eq.~(\ref{eq:bayes}), is strictly $0$. Non-zero probabilities can be
obtained only for bins including maximum citations greater than or
equal to the maximum value already obtained by this author.  (The
fact that the probabilities for these bins shown in
Fig.~\ref{fig:avePnms} are not strictly $0$ is a consequence of the
use of finite bin sizes.) Thus, binning authors on the basis of
their maximally cited paper {\em necessarily\/} underestimates their
quality. The mean, median and 65th percentile appear to be the most
balanced measures with roughly equal predictive value.

It is clear from Eq.~(\ref{eq:bayes}) that the ability of a given
measure to discriminate is greatest when the differences between the
conditional probability distributions, $P(i | \alpha )$, for
different author bins are largest. These differences can quantified
by measuring the `distance' between two such conditional
distributions with the aid of the Kullback-Leibler (KL) divergence
(also know as the relative entropy).  The KL divergence  between two
discrete probability distributions, $p$ and $p'$ is
defined\footnote{The non-standard choice of the natural logarithm
rather than the logarithm base two in the definition of the KL
divergence, will be justified below.} as
\begin{equation}\label{eq:kldef}
    \mathrm{KL}[p,p']=\sum_{i}p_i \ln \left(\frac{p_i}{p_i'}\right).
\end{equation}
The Kullback-Leibler divergence is positive and has desirable
convexity properties. It is, however, not a metric due to the fact
that $\mathrm{KL}[p',p] \ne \mathrm{KL}[p,p']$. While this asymmetry
is of little concern when the differences between $p$ and $p'$ are
small, some care is required when such differences are large.  This
can occur when the data set is so small that some citation bins are
empty or when we bin authors by $k_{\rm max}$, in which case empty
bins are inevitable as noted above.
\begin{figure}
\centering
  \includegraphics[width=\hsize]{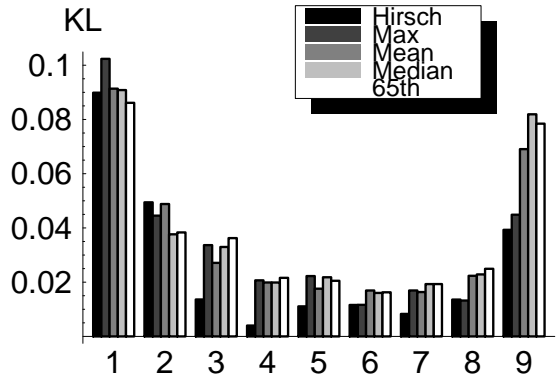}\\%
\caption{The Kullback-Leibler divergences
$\mathrm{KL}[P(i|\alpha),P(i|\alpha + 1)]$. Results are shown for
the following distributions:
$h$-index normalized by number of publications, maximum number of
citations, mean, median, and 65th percentile.
}\label{fig:KL}
\end{figure}
We consider the KL distance between adjacent distributions,
Fig.~\ref{fig:KL} shows the distances
$\mathrm{KL}[P(i|\alpha),P(i|\alpha+1)]$ for various measures. The
probability $P(\beta = \alpha + 1|\alpha)$ is exponentially
sensitive to the KL divergence. Measures with large KL divergences
between adjacent bins provide the most certain assignments of
authors. The KL divergences for the measures not shown are
significantly smaller than those displayed. The results of
Fig.~\ref{fig:KL} provide quantitative support for the roughly equal
performance of mean, median, and 65th percentile
measures\footnote{Figure~\ref{fig:KL} gives a misleading picture of
the $k_{\rm max}$ measure, since the KL divergences
$\mathrm{KL}[P(i|\alpha+1),P(i|\alpha)]$ are infinite as discussed
above.} seen in Figure~\ref{fig:avePnms}. The $h$-index normalized
by number of publications is dramatically smaller than the other
measures shown except for the extreme deciles.

The reduced ability of all measures to discriminate in the middle
deciles is immediately apparent from Fig.~\ref{fig:KL}.
\begin{figure}
  \includegraphics[width=\hsize]{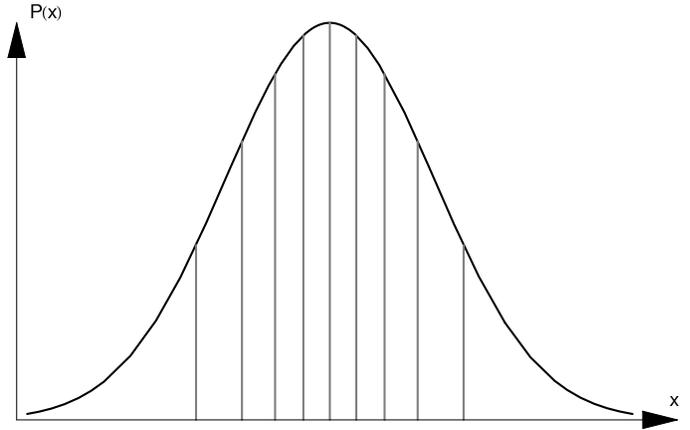}\\
  \caption{Binning according to deciles. This plot displays a normal distribution
  (solid black line) as an example of a probability  distribution
  peaked around a non-zero maximum. The grey vertical lines mark the
  boundaries of the $10$ deciles.
}\label{fig:decilebins}
\end{figure}
This is a direct consequence any percentile binning given that the
distribution of author quality has a maximum at some non-zero value,
the bin size of a percentile distribution near the maximum will
necessarily be small. The accuracy with which authors can be
assigned to a given bin in the region around the maximum is reduced
since one is attempting to distinguish between authors with very
similar citation distributions. As a result, the statistical
accuracy of percentile assignments is high at the extremes and
relatively low in the middle of the distribution where we are
attempting to make fine distinctions between scientists of similar
ability. This effect is illustrated in Fig.~\ref{fig:decilebins}.

\section{Scaling}\label{sec:scaling}
In this section, we consider the question of how many published
papers are required in order to make a reliable prediction of the
percentile ranking of a given author. (We consider results only
using the 65th percentile measure.) If this number is sufficiently
small, analysis along the lines presented here can provide a
practical tool of potential value in predicting long-term scientific
performance. In order to address this question, we will consider how
$P(\alpha|\{n_i\})$ scales as a function of the total number of
publications for an average author in each bin. Assume that an
average author belonging to bin $\alpha$ draws $N$ papers at random
from the distribution of $P(i |\alpha )$. The most probable number
of papers in each citation bin will thus be given as $n_i = N
P(i|\alpha)$. Inserting this result into Eq.~(\ref{eq:bayes}) and
discarding all fixed factors, we find that
\begin{equation}\label{eq:scaling2}
    P (\alpha |\{n_i\}) \sim p(\alpha )\left(\prod_{i}
    P(i|\alpha )^{P (i|\alpha )}\right)^N \ .
\end{equation}
For the same citation record, $\{ n_i \}$, a similar expression
permits determination of the probability that this average author
will be assigned to any bin, $\beta$.  We see that
\begin{equation}\label{eq:scaling3}
\lim_{N \to \infty} \, \frac{1}{N} \, \ln \left( \frac{P( \beta | \{
n_i \} )}{P( \alpha | \{ n_i \} )} \right) = - \mathrm{KL} \left[ P(
  \bullet |\alpha) , P(  \bullet  | \beta ) \right] \ .
\end{equation}
This equation illustrates the utility of the KL divergence and
explains the origin of its lack of symmetry.  It is clear from
Eqs.~(\ref{eq:scaling2}) and (\ref{eq:scaling3}) that the
probability of assigning this average author to the wrong bin will
ultimately vanish exponentially with $N$.  Given enough papers, the
largest bin will ultimately dominate.

To obtain a quantitative sense of how many papers are required in
practice, we pose the following question:  What is the probability
that a typical author
\begin{figure}
  \includegraphics[width=\hsize]{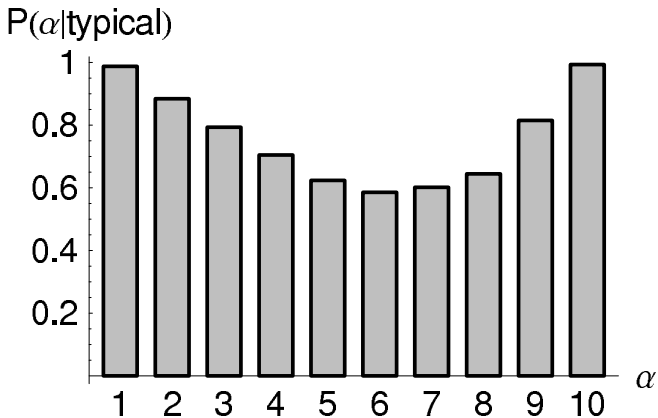}\\
\caption{The probability that a typical (i.e., most probable) author
with $50$ published papers will be assigned to the correct decile as
a function of actual author decile.  The median number of citations
is used as a measure.}\label{fig:scaling}
\end{figure}
from each author decile with $N=50$ published papers will be
assigned to the correct decile?  The answer is plotted as a
histogram in Fig.~\ref{fig:scaling} using the 65th percentile
citation rate as a measure (Similar results are obtained when using
the mean or median citation rates). The figure indicates that $N=50$
papers is more than sufficient to identify authors in the first and
tenth deciles. In fact, approximately $25$ and $20$ papers
respectively are sufficient to place authors in these deciles at the
90\% confidence level. Fig.~\ref{fig:scaling} also indicates that
$\approx 50$ published papers are sufficient to make meaningful
assignments of authors to the second, third, and ninth deciles.  All
measures have difficulty in assigning authors to deciles $5-8$. As
indicated by the small values of the KL divergence in these bins for
all measures considered, the citation distributions of these authors
are simply too similar to permit accurate discrimination (see
arguments in the previous section). On the other hand, the
probability that an author can be correctly assigned to one of these
middle bins on the basis of $50$ publication is high.  This
difficulty is due to the relatively small range of citations ranges
which cover these bins: the 65th percentile-bins 5 though 8 contain
authors with a 65th percentile between 5 and 13 citations (cf.~the
narrow ranges of the middle bins in the case of the mean, displayed
in Table~\ref{tab:authorbins}).

\section{Conclusions}\label{sec:conclusions}
There are two distinct questions which must be addressed in any
attempt to use citation data as an indication of author quality.
The first is whether the measure chosen to characterize a given
citation distribution or even the citation distribution itself
reflects the qualities that we would like to probe.  The second
question is whether a given measure is capable of discriminating
between authors in a statistically reliable way and, by extension,
which of several measures is best. We have shown that the use of
Bayesian statistics and the Kullback-Leibler divergence can answer
this question in a value-neutral and statistically compelling
manner.  It is possible to draw reliable conclusions regarding an
author's citation record on the basis of approximately 50 papers,
and it is possible to assign meaningful statistical uncertainties to
the results.  The high level of discrimination obtained in the
highest and lowest deciles provides indirect support for our
assumption that an author's citation record is drawn at random from
an appropriate conditional distribution and suggests that possible
additional correlations in citation data are not important.
Further, the difficulty in discriminating between authors in the
middle deciles suggests that intrinsic author ability is peaked at
some non-zero value.

The probabilistic methods adopted here permit meaningful comparison
of scientists working in distinct areas with only minimal value
judgments.  It seems fair, for example, to declare equality between
scientists in the same percentile of their peer groups.  It is
similarly possible to combine probabilities in order to assign a
meaningful ranking to authors with publications in several disjoint
areas.  All that is required is knowledge of the conditional
probabilities appropriate for each homogeneous subgroup.

We note, however, that the number of publications required to make
meaningful author assignments is large enough to limit the utility
of such analyses in the academic appointment process.  This raises
the question of whether there are more efficient measures of an
author's full citation record than those considered here.  Our
object has been to find that measure which is best able to assign
the most similar authors together.  Straightforward iterative
schemes can be constructed to this end and are found to converge
rapidly (i.e., exponentially) to an optimal binning of authors. (The
result is optimal in the sense that it maximizes the sum of the KL
divergences, $KL[P( \bullet  | \alpha), P( \bullet  | \beta)]$, over
all $\alpha$ and $\beta$.)  The results are only marginally better
than those obtained here with the mean, median or 65th percentile
measures.

Finally, it is also important to recognize that it takes time for a
paper to accumulate its full complement of citations.  While their
are indications that an author's early and late publications are
drawn (at random) on the same conditional distribution
\cite{lehmann:03}, many highly cited papers accumulate citations at
a constant rate for many years after their publication.  This
effect, which has not been addressed in the present analysis,
represents a serious limitation on the value of citation analyses
for younger authors.  The presence of this effect also poses the
additional question of whether there are other kinds of statistical
publication data that can deal with this problem. Co-author linkages
may provide a powerful supplement or alternative to citation data.
(Preliminary studies of the probability that authors in bins
$\alpha$ and $\beta$ will co-author a publication reveal a striking
concentration along the diagonal $\alpha = \beta$.) Since each paper
is created with its full set of co-authors, such information could
be useful in evaluating younger authors. This work will be reported
elsewhere.

\appendix
\section{Vertical Stripes}\label{sec:anomalies}
The most striking feature of the calculated $P( \beta | \alpha )$
shown in Fig.\,4 is presence of vertical `stripes'.  These stripes
are most pronounced for the poorest measures and disappear as the
reliability of the measure improves.  Here, we offer a schematic but
qualitatively reliable explanation of this phenomenon.  To this end,
imagine that each author's citation record is actually drawn at
random on the true distributions $Q(i|A)$.  For simplicity, assume
that every author has precisely $N$ publications, that each author
in true class $A$ has the same distribution of citations with
$n_i^{A} = N Q(i|A)$, and that there are equal numbers of authors in
each true author class.  These authors are then distributed into
author bins, $\alpha$, according to some chosen quality measure. The
methods of Sections IV and V can then be used to determine
$P(i|\alpha)$, $P( \{ n_i^{(A)} \} | \beta )$, $P( \beta | \{
n_i^{(A)} \})$ and $P (\beta | \alpha )$.  Given the form of the
$n_i^{(A)}$ and assuming that $N$ is large, we find that
\begin{equation}
P( \beta | \{ n_i^{(A)}\}) \approx \exp{\left( - N \mathrm{KL}[ Q( \bullet |A) , P(\bullet|\beta)] \right)}
\label{A1}
\end{equation}
and
\begin{equation}
\tilde{P}( \beta | \alpha) \sim \sum_A \, P(A | \alpha)
\exp{\left( - N \mathrm{KL}[ Q( \bullet |A) , P( \bullet |\beta)] \right)} \ ,
\label{A2}
\end{equation}
where $P(A | \alpha)$ is the probability that the citation record of
an author assigned to class $\alpha$ was actually drawn on $Q(i|A)$.
The results of this approximate evaluation are shown in Fig.\,8 and
compared with the exact values of $P( \beta | \alpha )$ for the
papers per year measure.  The approximations do not affect the
qualitative features of interest.

We now assume that the measure defining the author bins, $\alpha$,
provides a poor approximation to the true bins, $A$.  In this case,
authors will be roughly uniformly distributed, and the factor $P(A |
\alpha )$ appearing in Eq.\,(A2) will not show large variations.
Significant structure will arise from the exponential terms, where
the presence of the factor $N$ (assumed to be large), will amplify
the differences in the KL divergences. The KL divergence will have a
minimum value for some value of $A = A_0 (\beta)$, and this single
term will dominate the sum.  Thus, $\tilde{P}( \beta | \alpha )$
reduces to
\begin{equation}
\tilde{P}( \beta | \alpha) \sim P(A_0  | \alpha)
\exp{\left( - N \mathrm{KL}[ Q( \bullet |A_0 ) , P( \bullet |\beta)] \right)} \ .
\label{A3}
\end{equation}
The vertical stripes prominent in Figs.\,4(a) and (b) emerge as a
consequence of the dominant $\beta$-dependent exponential factor.
\begin{figure}
 \begin{center}
  \begin{tabular}{cc}
    \emph{(a)} Papers/year $\tilde{P}(\alpha'|\alpha)$ & \emph{(b)} Papers/year $P(\alpha'|\alpha)$  \\
    \includegraphics[width=.48\hsize]{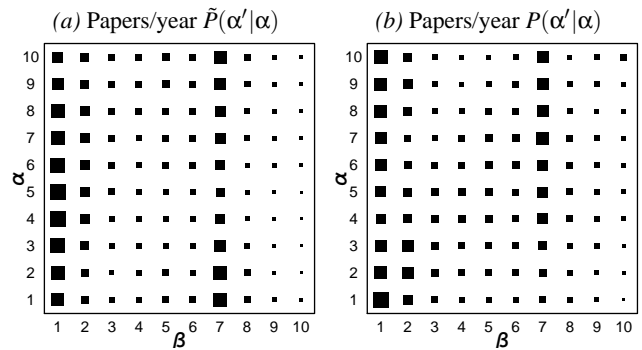} &
    \includegraphics[width=.48\hsize]{f4_ppyPmn}
  \end{tabular}
  \caption{A comparison of the approximate $\tilde{P}(\beta|\alpha)$  from
  Eq.~(\ref{A2}) and the exact $P(\beta|\alpha)$ for the papers published
  per year measure.}\label{fig:ppyhirsch}
 \end{center}
\end{figure}
The present arguments also apply to the worst possible measure,
i.e., a completely random assignment of authors to the bins
$\alpha$.  In the limit of a large number of authors, $N_{\rm aut}$,
all $P(i | \beta )$ will be equal except for statistical
fluctuations.  The resulting KL divergences will respond linearly to
these fluctuations.\footnote{This is true because there will be no
choice of $A$ such that $Q(i |A) = P(i | \alpha)$.}  These
fluctuations will be amplified as before provided only that $N_{\rm
aut}$ grows less rapidly than $N^2$.  The argument here does {\em
not\/} apply to good measures where there is significant structure
in the term $P(A | \alpha)$.  (For a perfect measure, $P(A | \alpha)
= \delta_{A \alpha}$.)  In the case of good measures, the expected
dominance of diagonal terms (seen in the lower row of Fig.\,4)
remains unchallenged.

\section{Explicit Distributions}\label{sec:explicit}
For convenience we present all data to determine the probabilities
$P(\alpha|\{n_i\})$ for authors who publish in the theory
sub-section of SPIRES. Data is presented only for case of the mean
number of citations. All citations are binned logarithmically
according to the citation bins listed in column one and two of
Table~\ref{tab:binning}.
\begin{table}
  \centering
  \begin{tabular}{|c|r@{\,$k$\,}l||c|r@{\,$N$\,}l|}
    \hline
    \multicolumn{3}{|c||}{$P(i|\alpha)$} & \multicolumn{3}{|c|}{$P(N|\alpha)$}\\\hline
    Bin number & \multicolumn{2}{c||}{Citation range} & Bin Number & \multicolumn{2}{c|}{Total paper range} \\ \hline
    $i=1$ &       & $= 1$        & $m=1$ &        &$= 25$ \\
    $i=2$ &       & $= 2$        & $m=2$ &        &$= 26$ \\
    $i=3$ &    $2< $&$ \leq 4$   & $m=3$ & $26 < $&$ \leq 28$ \\
    $i=4$ &    $4< $&$ \leq 8$   & $m=4$ & $28 < $&$ \leq 32$ \\
    $i=5$ &    $8< $&$ \leq 16$  & $m=5$ & $32 < $&$ \leq 40$ \\
    $i=6$ &   $16< $&$ \leq 32$  & $m=6$ & $40 < $&$ \leq 56$ \\
    $i=7$ &   $32< $&$ \leq 64$  & $m=7$ & $56 < $&$ \leq 88$ \\
    $i=8$ &   $64< $&$ \leq 128$ & $m=8$ & $88 < $&$ \leq 152\,$ \\
    $i=9$ &  $128< $&$ \leq 256$ & $m=9$ & $\,152< $&$ \leq N_{\rm max}$ \\\cline{4-6}
    $i=10$ & $256< $&$ \leq 512\,$ &  \multicolumn{3}{c|}{}\\
    $i=11$ & $\,512< $&$ \leq k_{\rm max}$ & \multicolumn{3}{c|}{} \\
    \hline
  \end{tabular}
  \caption{The binning of citations and total number of papers.
  The first and second column show the bin number and bin ranges
  for the citation bins used to determine the conditional citation probabilities
  $P(i|\alpha)$ for each $\alpha$, shown in Table~\ref{tab:Pialpha}.
  The third and fourth
  column display the bin number and total number of paper ranges used in the
  creation of the conditional probabilities $P(m|\alpha)$ for each $\alpha$,
  displayed in Table~\ref{tab:PNalpha}.}\label{tab:binning}
\end{table}
The author bins are determined on the basis of deciles of the total
distribution of mean citations, $p(\langle k \rangle)$.
Table~\ref{tab:authorbins} shows the relevant quantities for these
bins.
\begin{table}
  \centering
  \begin{tabular}{|c|c@{--}c|c|c|c|}
    \hline
    $\alpha$ & \multicolumn{2}{c|}{$\langle k \rangle$-range} & \# authors & $p(\alpha)$ & $\bar{n}(\alpha)$\\\hline
    1 & $0 $&$ 1.69$ & 673     & 0.1 & 37.0\\
    2 & $1.69 $&$ 3.08$ & 673  & 0.1 & 41.8\\
    3 & $3.08 $&$ 4.88$ & 675  & 0.1 & 44.0\\
    4 & $4.88 $&$ 6.94$ & 673  & 0.1 & 46.8\\
    5 & $6.94 $&$ 9.40$ & 674  & 0.1 & 52.2\\
    6 & $9.40 $&$ 12.56$ & 674 & 0.1 & 54.3\\
    7 & $12.56 $&$ 16.63$ & 673& 0.1 & 59.5\\
    8 & $16.63 $&$ 22.19$ & 674& 0.1 & 59.0\\
    9 & $22.19 $&$ 33.99$ &674 & 0.1 & 65.4\\
    10 & $\,\,33.99 $&$ 285.88\,\,$ &674& 0.1& 72.2\\
    \hline
  \end{tabular}
  \caption{The author bins. This table shows the mean numbers of citations
  that define the limits of the $10$ author bins.}\label{tab:authorbins}
\end{table}
Given the definitions of both the author- and citation bins, we can
determine the conditional citation distributions $P(i|\alpha)$
empirically. These are given in Table~\ref{tab:Pialpha}.
\begin{table*}
  \centering
  \begin{tabular}{|c|ccccccccccc|}\hline
   & $i = 1$ & $i = 2$ & $i = 3$ & $i = 4$ & $i = 5$ & $i = 6$ & $i = 7$ & $i = 8$ & $i = 9$ & $i = 10$ & $i = 11$\\\hline
 $\alpha = 1$ & 0.612 & 0.182 & 0.127 & 0.057 & 0.019 & 0.002 & 0.000 & 0.000 & 0.000 & 0.000 & 0.000 \\
 $\alpha = 2$ & 0.433 & 0.188 & 0.181 & 0.122 & 0.055 & 0.016 & 0.004 & 0.000 & 0.000 & 0.000 & 0.000 \\
 $\alpha = 3$ & 0.327 & 0.165 & 0.188 & 0.167 & 0.103 & 0.038 & 0.010 & 0.002 & 0.000 & 0.000 & 0.000 \\
 $\alpha = 4$ & 0.263 & 0.143 & 0.178 & 0.184 & 0.140 & 0.067 & 0.019 & 0.005 & 0.001 & 0.000 & 0.000 \\
 $\alpha = 5$ & 0.217 & 0.127 & 0.163 & 0.183 & 0.165 & 0.096 & 0.036 & 0.009 & 0.002 & 0.000 & 0.000 \\
 $\alpha = 6$ & 0.177 & 0.113 & 0.150 & 0.181 & 0.173 & 0.126 & 0.058 & 0.017 & 0.004 & 0.001 & 0.000 \\
 $\alpha = 7$ & 0.143 & 0.098 & 0.135 & 0.170 & 0.183 & 0.149 & 0.086 & 0.028 & 0.007 & 0.002 & 0.000 \\
 $\alpha = 8$ & 0.118 & 0.080 & 0.121 & 0.155 & 0.182 & 0.169 & 0.110 & 0.048 & 0.012 & 0.003 & 0.000 \\
 $\alpha = 9$ & 0.094 & 0.066 & 0.099 & 0.141 & 0.175 & 0.178 & 0.139 & 0.075 & 0.025 & 0.007 & 0.001 \\
 $\alpha = 10$ & 0.068 & 0.045 & 0.071 & 0.107 & 0.145 & 0.171 & 0.166 & 0.121 & 0.067 & 0.027 & 0.012  \\\hline
\end{tabular}
  \caption{The distributions $P(i|\alpha)$. This table
  displays the conditional probabilities that an author writes a paper in paper-bin $i$
  given that his author-bin is $\alpha$.}\label{tab:Pialpha}
\end{table*}

We also need the probabilities $P(N|\alpha)$ describing that an
author in bin $\alpha$ has $N$ publications. Because of the low
number of authors in each bin, we need to bin the total number of
publications when calculating this probability; we use the letter
$m$ to enumerate the $N$-bins. Because $P(N|\alpha)$ is described by
a power-law distribution\footnote{This fact is known as
\textit{Lotka's Law} \cite{lotka:26}.} and since we only consider
authors with more than $25$ publications, we choose to bin $N$
logarithmically as displayed in the third and fourth column of
Table~\ref{tab:binning}. The conditional probabilities,
$P(m|\alpha)$ are displayed in Table~\ref{tab:PNalpha}.
\begin{table*}
  \centering
  \begin{tabular}{|c|ccccccccc|}\hline
     & $m=1$ & $m=2$ & $m=3$ & $m=4$ & $m=5$ & $m=6$ & $m=7$ & $m=8$ & $m=9$ \\\hline
   $\alpha = 1$ & 0.083 & 0.071 & 0.134 & 0.226 & 0.224 & 0.172 & 0.082 & 0.006 & 0.001 \\
   $\alpha = 2$ & 0.058 & 0.049 & 0.103 & 0.187 & 0.236 & 0.217 & 0.122 & 0.025 & 0.003 \\
   $\alpha = 3$ & 0.068 & 0.050 & 0.095 & 0.133 & 0.231 & 0.240 & 0.136 & 0.041 & 0.004 \\
   $\alpha = 4$ & 0.043 & 0.049 & 0.095 & 0.141 & 0.198 & 0.247 & 0.162 & 0.061 & 0.004 \\
   $\alpha = 5$ & 0.031 & 0.059 & 0.067 & 0.108 & 0.181 & 0.246 & 0.200 & 0.091 & 0.016 \\
   $\alpha = 6$ & 0.031 & 0.039 & 0.068 & 0.126 & 0.162 & 0.245 & 0.215 & 0.099 & 0.015 \\
   $\alpha = 7$ & 0.034 & 0.022 & 0.058 & 0.114 & 0.152 & 0.242 & 0.215 & 0.128 & 0.034 \\
   $\alpha = 8$ & 0.028 & 0.024 & 0.049 & 0.096 & 0.178 & 0.243 & 0.248 & 0.101 & 0.033 \\
   $\alpha = 9$ & 0.030 & 0.033 & 0.037 & 0.074 & 0.148 & 0.228 & 0.245 & 0.160 & 0.045 \\
  $\alpha = 10$ & 0.027 & 0.028 & 0.043 & 0.077 & 0.131 & 0.212 & 0.199 & 0.223 & 0.061\\\hline
  \end{tabular}
  \caption{The conditional probabilities $P(m|\alpha)$. This table contains the
  conditional probabilities that an author has a total
  number of publications in publication-bin $m$ given that his author-bin is
  $\alpha$ .}\label{tab:PNalpha}
\end{table*}


\end{document}